\documentclass[9pt]{elife}
\usepackage{lipsum} 
\usepackage[version=4]{mhchem}
\usepackage{siunitx}
\DeclareSIUnit\Molar{M}

\newcommand{\lm}[1]{\textcolor{black}{{#1}}} %

\title{Neural mechanisms underlying the temporal organization of naturalistic animal behavior}

\author[1*]{Luca Mazzucato}
\affil{Institute of Neuroscience, Departments of Biology, Mathematics and Physics, University of Oregon, Eugene.}

\corr{lmazzuca@uoregon.edu}{FMS}

\begin{document}

\maketitle

\begin{abstract}
Naturalistic animal behavior exhibits a strikingly complex organization in the temporal domain, whose variability stems from at \lm{least three sources: hierarchical, contextual, and stochastic}. What are the neural mechanisms and computational principles generating such complex temporal features? In this review, we provide a \lm{critical assessment of} the existing behavioral and neurophysiological evidence for these sources of temporal \lm{variability} in naturalistic behavior. We crystallize recent studies which converge on an emergent mechanistic theory of temporal variability based on attractor neural networks and metastable dynamics, arising from the coordinated interactions between mesoscopic neural circuits. We highlight the crucial role played by structural heterogeneities and by noise arising in mesoscopic circuits. We assess the shortcomings and missing links in the current \lm{theoretical and experimental} literature and propose new directions of investigations \lm{to fill these gaps}.
\end{abstract}

\section{Introduction}
\label{sec:intro}

\begin{figure}[t]
\includegraphics[width=\linewidth]{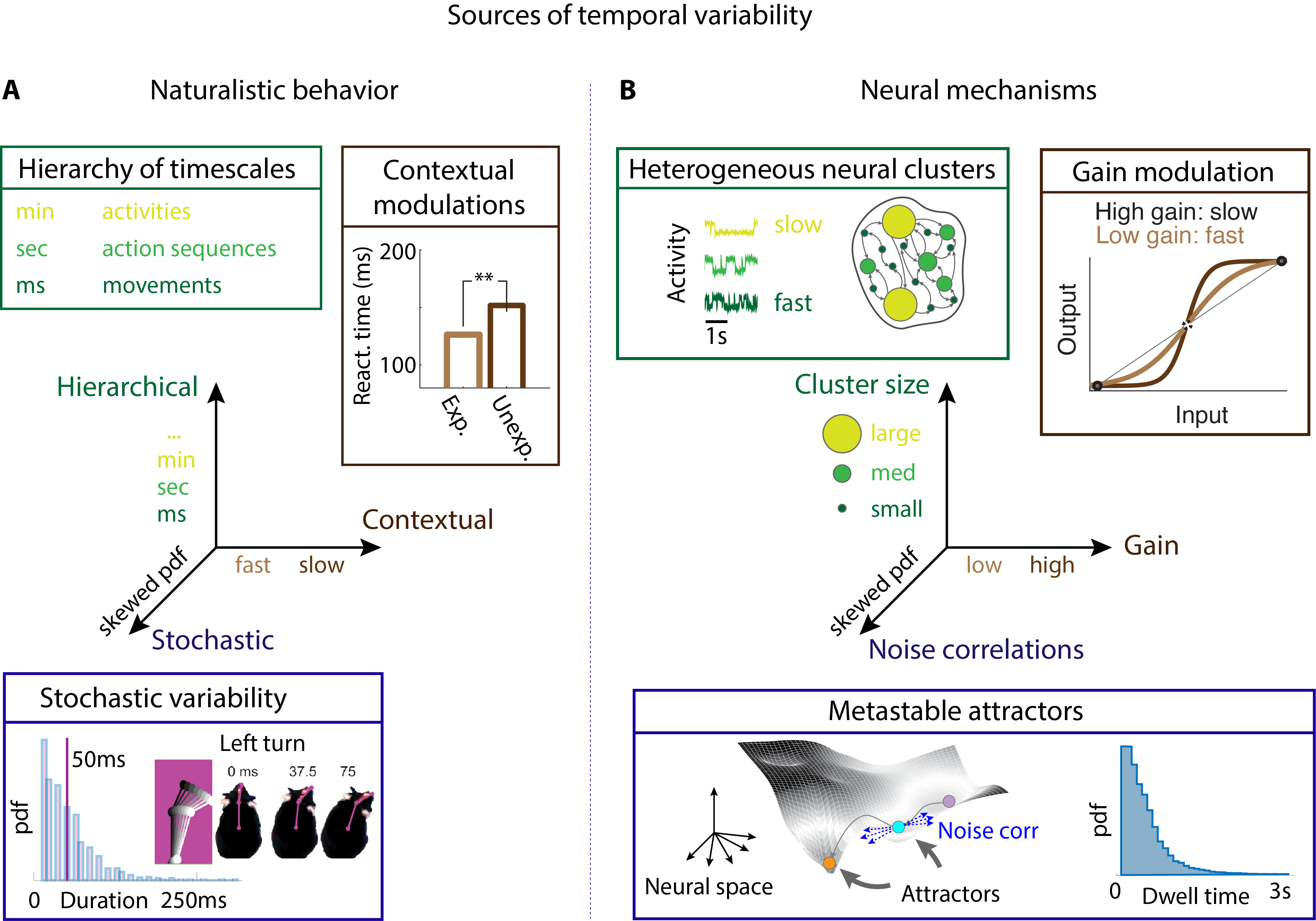}
\caption{{\it Neural mechanisms underlying the temporal organization of naturalistic animal behavior.} {\bf A)} Three sources of temporal variability in naturalistic behavior: hierarchical (from fast movements, to behavioral action sequences, to slow activities and long-term goals), contextual (reaction times are faster when stimuli are expected), and \lm{stochastic} (the distribution of "turn right" action in freely moving rats is right-skewed). {\bf B)} Neural mechanisms underlying each source of temporal variability: hierarchical variability may arise from recurrent networks with a heterogeneous distribution of neural cluster sizes; contextual modulations from neuronal gain modulation; stochastic variability from metastable attractor dynamics where transitions between attractors are driven by low-dimensional noise, leading to right-skewed distributions of attractor dwell times. Panel A adapted from Fig. 2  \citep{jaramillo2011auditory}, with permission from Elsevier. It is covered by the CC-BY 4.0 license and further reproduction of this panel would need permission from the copyright holder. Panel A, bottom, reproduced from Fig. 6 \citep{findley2021sniff}.}
\label{figone}
\end{figure}

Naturalistic animal behavior exhibits a striking amount of variability in the temporal domain (Fig. \ref{figone}A). \lm{An individual animal's} behavioral variability can be decomposed across at least three axes: hierarchical, contextual, and \lm{stochastic}. The first source of variability originates from the vast hierarchy of timescales underlying self-initiated, spontaneous behavior ranging from milliseconds to minutes in animals (and to years for humans). On the sub-second timescale, animals perform fast movements varying from tens to hundreds of milliseconds. In rodents, these movements include whisking, sniffing, and moving their limbs. On the slower timescales of seconds, animals concatenate these fast movements into behavioral sequences of self-initiated actions, such as exploratory sequences (moving around an object while sniffing, whisking and wagging their noses) or locomotion sequences (coordinating limb and head movements to reach a landmark). These sequences follow specific syntax rules \citep{berridge1987natural} and can last several seconds. On the timescales of minutes or longer, mice may repeat the "walk and explore" behavioral sequence multiple times, when engaged in some specific activity, such as foraging, persisting towards their goal for long periods. In this simple example, a freely moving mouse exhibits behavior whose temporal organization vary over several orders of magnitudes simultaneously, ranging from the sub-second scale (actions), to several seconds (behavioral sequences), to minutes (goals to attain). A leading theory to explain the temporal organization of naturalistic behavior is that behavioral action sequences arrange in a hierarchical structure \citep{tinbergen2020study,dawkins1976hierarchical,simon1991architecture}, where actions are nested into behavioral sequences which are then grouped into activities. This hierarchy of timescales is ubiquitously observed across species during naturalistic behavior.

\lm{The second source of temporal variability stems from the different contexts in which a specific behavior can be performed. For example, reaction times to a delivered stimulus can be consistently faster when an animals is expecting it, compared to slower reaction times when the stimulus is unexpected. Contextual sources can be either internally generated (modulations of brain state, arousal, expectation), or externally driven (pharmacological, optogenetic or genetic manipulations; or changes in task difficulty or environment). }

\lm{After controlling for all known contextual effects, an individual's behavior still exhibits a large amount of residual temporal variability across repetitions of the same behavioral unit. This residual variability across repetitions can be quantified as a property of the distribution of movement durations, which is typically very skewed with an exponential tail \citep{wiltschko2015mapping,findley2021sniff,recanatesi2022metastable}, suggesting a stochastic origin. We define as stochastic variability the fraction of variability in the expression of a particular behavioral unit, which cannot be explained by any readily measurable variables.}


\lm{These three aspects of temporal variability (hierarchical, contextual, stochastic) can all be observed in the naturalistic behavior of a single individual during their lifetime. One additional source of behavioral variability is individual phenotypic variability across different animals, an important aspect of behavior in ethological and evolutionary light \citep{honegger2018stochasticity}. Although we will briefly discuss some of its features below, the main focus of this review will be on the three main axes of variability described above, which pertain the behavior of a single animal.}

From this overview we conclude that the temporal structure of naturalistic, self-initiated behavior can be decomposed along at least three axes of temporal variability (Fig. \ref{figone}a): \lm{a large hierarchy of simultaneous timescales at which behavior unfolds, ranging from milliseconds to minutes; contextual modulations affecting the expression of behavior at each level of this hierarchy; and stochasticity in the variability of the same behavioral unit across repetition in the same context.} These three axes may or may not interact depending on the scenarios and the species. Here, we will review recent theoretical and computational results establishing the foundations of a mechanistic theory that explains how these three sources of temporal variability can arise from biologically plausible computational motifs. More specifically, we will address the following questions: 
\begin{itemize}
\item {\bf Neural representation:} How are self-initiated actions represented in the brain and how are they concatenated into behavioral sequences? 
\item {\bf \lm{Stochasticity}:} What neural mechanisms underlie the temporal variability observed in behavioral units \lm{across multiple repetitions within the same context}? 
\item {\bf Context:} How do contextual modulations affect the temporal variability in behavior, enabling flexibility in action timing and behavioral sequence structure? 
\item {\bf Hierarchy:} How do neural circuits generate the vast hierarchy of timescales from milliseconds to minutes, hallmark of naturalistic behavior?
\end{itemize}
The mechanistic approach we will review is based on the theory of metastable attractors (Fig. \ref{figone}b), which is emerging as a unifying principle expounding many different aspects of the dynamics and function of neural circuits \citep{la2019cortical}. We will first establish a precise correspondence between behavioral units and neural attractors at the level of self-initiated actions (the lowest level of the temporal hierarchy). Then we will show how the emergence of behavioral sequences originates from sequences of metastable attractors. Our starting point is the observation that transitions between metastable attractors can be driven by the neural variability internally generated within a local recurrent circuit. This mechanism can naturally explain the action timing variability \lm{of stochastic origin. We will examine biologically plausible mesoscopic circuits which can learn to flexibly execute complex behavioral sequences}. We will then review the neural mechanisms underlying contextual modulations of behavioral variability. We will show that the average transition time between metastable attractors can be regulated by changes in single-cell gain. Gain modulation is a principled neural mechanism mediating the effects of context, which can be induced by either internal or external perturbations and supported by different neuromodulatory and cortico-cortical pathways; or by external pharmacological or experimental interventions. Finally, we will review how a large hierarchy of timescales can naturally and robustly emerge from heterogeneities in a circuit's structural connectivity motifs, such as neural clusters with heterogeneous sizes. Although most of the review is focused on the behavior of \lm{individual} animals, we discuss how recent results on multi-animal interactions and social behavior may challenge existing theories of naturalistic behavior and brain function. \lm{Three Appendices provide guides to computational methods for behavioral video analyses and modeling; theoretical and experimental aspects of attractors dynamics; and biologically plausible models of metastable attractors.}

\begin{table}[bt]
\caption{\label{box:definitions}Definitions of behavioral units \lm{(see Appendix \ref{box:videos} for how to extract these features from videos).}}
\begin{tabular}{|p{1.7cm}|p{11.7cm}|}
\hline
Action & The simplest building blocks of behavior at the lowest level of the hierarchy, which occur at a fast, typically sub-second timescale, and cannot be further decomposed into smaller units. \lm{We define an action as a short stereotypical trajectory in posture space \citep{brown2018ethology} (synonyms include movemes, syllables \citep{anderson2014toward}).} Examples include poking in or poking out of a nose port, waiting at a port, pressing a lever. A widely used operational definition identifies actions as the discrete latent \lm{trajectories of an autoregressive state space model} fit to pose-tracking time series data (Fig. \ref{figtwo}) \citep{wiltschko2015mapping,findley2021sniff}. \lm{An alternative definition is in terms of short spectrotemporal representations from a time-frequency analysis of videos \citep{berman2016predictability,marshall2021continuous}.} \\\hline 
Behavioral action sequence &  A combination of actions concatenated in a meaningful yet stereotyped way, lasting up to a few seconds. A sequence can occur during trial-based experimental protocols (e.g.: a short sequence of actions aimed at obtaining a reward in an operant conditioning task \citep{geddes2018optogenetic,murakami2014neural}; running between opposite ends of a linear track \citep{maboudi2018uncovering}) or during spontaneous periods (e.g.: repeatedly scratching own's head; picking up and manipulating an object). \\\hline
Activity & A concatenation of multiple behavioral sequences, often repeated and of variable duration, typically aimed at obtaining a goal and lasting up to minutes or even hours. Examples include grooming, foraging, mating, feeding, exploration. Activities typically unfold in naturalistic freely moving settings devoid of experimenter-controlled trial structure. \\\hline
\end{tabular}
\end{table}

\section{\lm{The stochastic nature of naturalistic behavior}}

\begin{figure}[t]
\begin{fullwidth}
\includegraphics[width=\linewidth]{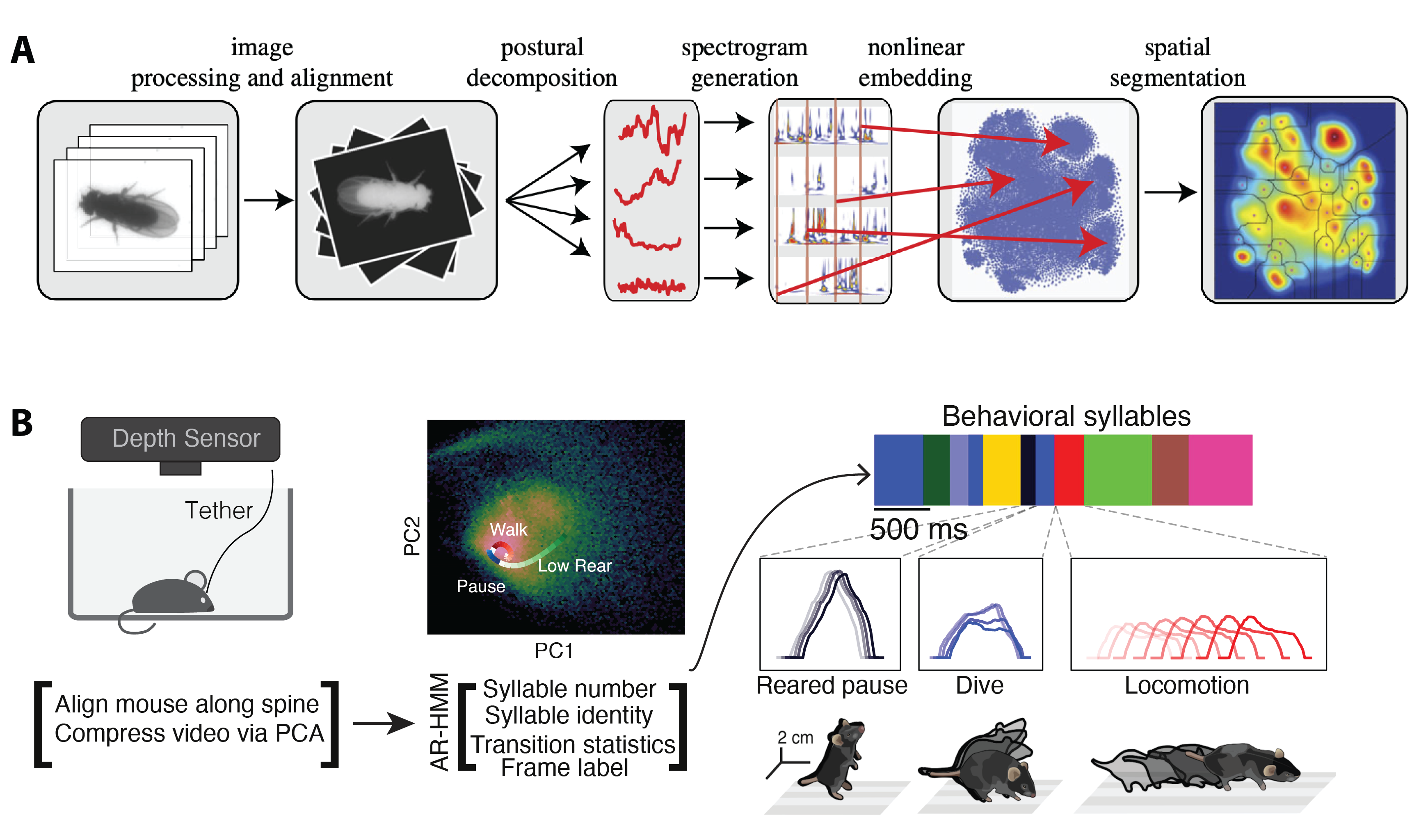}
\caption{{\it The complex spatiotemporal structure in naturalistic behavior.} {\bf A)} Identification of behavioral syllables in the fruit fly. From left to right: Raw images of {\it Drosophila} melanogaster (1) are segmented from background, rescaled and aligned (2), then decomposed via PCA into a low-dimensional time series (3). A Morlet wavelet transform yields a spectrogram for each postural mode (4), mapped into a two-dimensional plane via t-SNE (5). A watershed transform \lm{identifies} individual peaks from one another (6). {\bf B)} Identification of behavioral action sequences from 3D videos with MoSeq. An autoregressive Hidden Markov Model (AR-HMM, right) fit to PCA-based video compression (center) identifies hidden states representing actions (color bars, right, top: color-coded intervals where each HMM state is detected). Panel A reproduced from Fig. 2 and 5 \citep{berman2014mapping}, with permission from Royal Society Publishing. Panel B reproduced from Fig. 1 \citep{wiltschko2015mapping}, with permission from Elsevier. They are not covered by the CC-BY 4.0 license and further reproduction of this panel would need permission from the copyright holder.}
\label{figtwo}
\end{fullwidth}
\end{figure}

The study of naturalistic behavior based on animal videos has recently undergone a revolution due to the spectacular accuracy and efficiency of \lm{computational methods} for animal pose tracking \citep{stephens2008dimensionality,berman2014mapping,hong2015automated,mathis2018deeplabcut,pereira2019fast,nourizonoz2020etholoop,segalin2020mouse} \lm{(see Appendix \ref{box:videos} and Table \ref{box:definitions} for details)}. These new methods work across species and conditions, ushering a new era for computational neuroethology \citep{anderson2014toward,brown2018ethology,datta2019computational}. They have led to uncovering a quantitive classification of self-initiated behavior revealing a stunning amount of variability both in its lexical features (which actions to choose, in which order) and in its temporal dimension (when to act) \citep{berman2016predictability,wiltschko2015mapping,markowitz2018striatum,marshall2021continuous,recanatesi2022metastable}. \lm{Different ways to characterize the behavioral repertoire on short timescales have been developed, depending on the underlying assumptions of whether the basic units of behavior are discrete or continuous (Appendix \ref{box:videos}). One can define stereotypical behaviors or postures by clustering probability density maps of spectro-temporal features extracted from behavioral videos, as in the example of fruit flies \citep{berman2014mapping,berman2016predictability} or mice \citep{marshall2021continuous} (Fig. \ref{figtwo}A). Alternatively, one can capture actions or postures as discrete states of a state space model based on Markovian dynamics, each state represented as a latent state autoregressive trajectory accounting for \lm{stochastic} movement variability (Fig. \ref{figtwo}B). State space models were applied successfully to C. elegans \citep{linderman2019hierarchical}, {\it Drosophila} \citep{tao2019statistical} (although heavy-tailed distributions were reported in tethered flies \citep{maye2007order}), zebrafish \citep{johnson2020probabilistic} and rodents \citep{wiltschko2015mapping,markowitz2018striatum}. At the shortest timescale of actions and postures, transition times between consecutive actions are well described by a Poisson process \citep{killeen1988behavioral}, characterized by a right-skewed distribution of inter-action intervals (Fig. \ref{figthree}). In this and the next section, we will focus on short timescales (up to a few seconds) where state space models provide parsimonious accounts of behavior. However, these models fail to account for longer timescale and non-Markovian structure in behavior, and in later sections we will move to a more data-driven approach to investigate these aspects of behavior (see Appendix \ref{box:videos}).}

\subsection{Self-initiated actions and ensemble activity patterns in premotor areas}

\begin{figure}[t]
\begin{fullwidth}
\includegraphics[width=\linewidth]{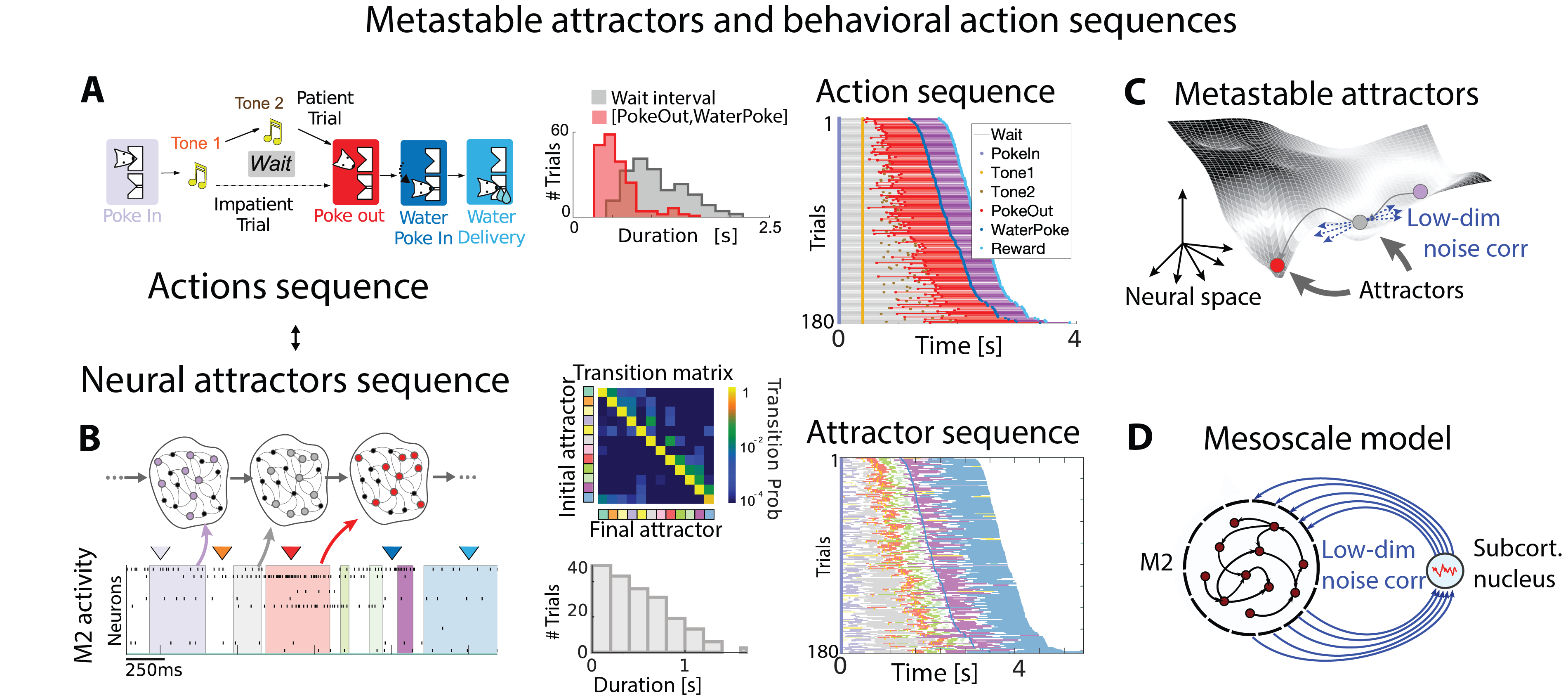}
\caption{{\it Metastable attractors in secondary motor cortex can account for the stochasticity in action timing.} {\bf A} In the Marshmallow experiment, a freely moving rat \lm{poked into a wait port, where tone 1 signaled the trial start. The rat could poke out at any time (after tone 2, played at random times, in patient trials; or before tone 2, in impatient trials) and move to the reward port to receive a water reward (large and small for patient and impatient trials, respectively).} Center: Distributions of wait times (gray) and intervals between Poke Out and Water Poke (red) reveal a large temporal variability across trials. Right: Behavioral sequences (sorted from shortest to longest) in a representative session. {\bf B} Left: Representative neural ensemble activity from M2 during an impatient trial (tick marks indicate spikes; colored arrows indicate the rat's actions, \lm{same colors as in A}) \lm{with overlaid hidden Markov model states, interpreted as neural attractors each represented as a set of coactivated neurons within a network (colored intervals indicate HMM states detected with probability above 80\%)}. Top center: Transition probability matrix between HMM states. Bottom center: The distribution of state durations (representative gray pattern from left plot) is right-skewed, suggesting a stochastic origin of state transitions. Right: Sequence of color-coded HMM states from all trials in the representative session of panel A. {\bf C} Schematic of an attractor landscape: attractors representing HMM states in panel E are shown as potential wells. Transitions between consecutive attractors are driven by low-dimensional correlated noise. {\bf D} Schematic of a mesoscopic neural circuit generating stable attractor sequences with variable transition times comprising a feedback loop between M2 and a subcortical nucleus. Reproduced from Fig. 1 and 5 \citep{recanatesi2022metastable}, with permission from Elsevier. They are not covered by the CC-BY 4.0 license and further reproduction of this panel would need permission from the copyright holder.
}
\label{figthree}
\end{fullwidth}
\end{figure}

These foundational studies revealed a vast repertoire of \lm{dozens to} hundreds of actions \citep{berman2014mapping,wiltschko2015mapping,markowitz2018striatum,marques2018structure,costa2019adaptive,schwarz2015changes} \lm{(although the repertoire depends on the coarse-graining scale of the behavioral analysis)}, leading to a combinatorial explosion in the number of possible action sequences. To tame the curse of dimensionality, typical of unconstrained naturalistic behavior, a promising approach is to control for the lexical variability in behavioral sequences and design a naturalistic task with a single behavioral sequence, though retaining its temporal variability. The authors of  \citep{murakami2014neural,murakami2017distinct,recanatesi2022metastable} adopted this strategy and to train freely moving rats to perform a self-initiated task (the rodent version of the "Marshmallow task,"  \citep{mischel1970attention}, Fig. \ref{figthree}A) where a specific set of actions had to be performed in a fixed order to obtain a reward. The many repetitions of the same behavioral sequence yielded a large sample size to elucidate the source of temporal variability across trials. Action timing retained the temporal variability hallmark of naturalistic behavior, characterized by skewed distributions of action durations (Fig. \ref{figthree}A).

What is the neural mechanism generating the large variability in action timing? A large number of studies implicated the secondary motor cortex (M2) in rodents as part of a distributed network involved in motor planning in head-fixed mice \citep{li2016robust,barthas2017secondary} and controlling the timing of self-initiated actions in freely moving rats \citep{murakami2014neural,murakami2017distinct}. 
\lm{Ensemble activity recorded in M2 during the Marshmallow task unfolded via sequences of multi-neuron firing patterns, each one lasting hundreds of milliseconds to a few seconds; within each pattern, neurons fired at an approximately constant rate (Fig. \ref{figthree}B).} Such long dwell times, much longer than typical single neuron time constants, suggest that the observed metastable patterns may be an emergent property of the collective circuit dynamics within M2 and reciprocally connected brain regions. Crucially, both neural and behavioral sequences were highly reliable yet temporally variable, and the distribution durations of action and neural pattern durations were characterized by a right-skewed distribution. This temporal heterogeneity suggests that a stochastic mechanism, such as found in noise-driven transitions between metastable states, could contribute to driving transitions between consecutive patterns within a sequence (see below). \lm{A dictionary between actions and neural pattern could be established, revealing that the onset of specific patterns} reliably preceded upcoming self-initiated actions (Fig. \ref{figthree}A-B, \lm{e.g., the onset of the red pattern reliably precedes the poke out movement}). The dictionary trained on the rewarded sequence generalized to epochs where the animal performed erratic non-rewarded behavior, where attractor onset predicted upcoming actions as well.
\lm{The use of state space models with underlying Markovian dynamics as generative models, capturing both naturalistic behavior \citep{wiltschko2015mapping,batty2019behavenet,johnson2020probabilistic,2021distance,findley2021sniff} and the underlying neural pattern sequences \citep{maboudi2018uncovering,linderman2019hierarchical,recanatesi2022metastable} is a powerful tool to bridge the first two levels of the temporal hierarchy in naturalistic behavior: a link from actions to behavioral sequences.} This generative framework further revealed fundamental aspects of neural coding in M2 ensembles, such as their distributed representations, dense coding and single-cell multistability \citep{recanatesi2022metastable,mazzucato2015dynamics}.

\subsection{Action timing variability from metastable attractors}

Neural patterns in M2 may represent attractors of the underlying recurrent circuit dynamics. An attractor is a persisting pattern of population activity where neurons fire at an approximately constant rate for an extended period of time (see Appendix \ref{box:attractors} for models of attractor dynamics). Foundational work in head-fixed mice showed that licking preparatory activity in M2 during a delayed response task is encoded by choice-specific discrete attractors \citep{inagaki2019discrete}. Attractors can be stable \citep{amit1997model}, as observed in monkey IT cortex in working memory tasks \citep{fuster1981inferotemporal,miyashita1988neuronal}. Attractors can also be metastable (see Appendix \ref{box:metaattractors}), when they typically last for hundreds of milliseconds and noise fluctuations spontaneously trigger transitions to a different attractor \citep{deco2012neural,LitwinKumarDoiron2012}. 

Metastable attractors can be concatenated into sequences, which can either be random, as observed during ongoing periods in sensory cortex \citep{mazzucato2015dynamics}, or highly reliable, encoding the evoked response to specific sensory stimuli \citep{Jones2007,miller2010stochastic,mazzucato2015dynamics}, or underlying freely moving behavior \citep{maboudi2018uncovering,recanatesi2022metastable}. In particular, M2 ensemble activity in the Marshmallow experiment was consistent with the activity generated by a specific sequence of metastable attractors (Fig. \ref{figthree}). \lm{The main hypothesis underlying this model is that the onset of an attractor drives the initiation of a specific action as determined by the action/pattern dictionary (Fig. \ref{figthree}A-B) and the dwell-time in a given attractor sets the inter-action-interval. The dynamics of the relevant motor output and the details of variability in movement execution is generated downstream of this attractor circuit (see Fig. \ref{figfive} and Discussion section).} The main features observed in the M2 ensemble dynamics during the Marshmallow task (i.e., long-lived patterns with a right-skewed dwell-time distribution, concatenated into highly reliable pattern sequences) can be explained by a two-area mesoscopic network where a large recurrent circuit (representing M2) is reciprocally connected to a small circuit lacking recurrent couplings (a subcortical area likely representing the thalamus \citep{guo2018anterolateral,guo2017maintenance} or the basal ganglia \citep{helie2015learning,desmurget2010motor,nakajima2019prefrontal}, see Fig. \ref{figfive}B). In this biologically plausible model, metastable attractors are encoded in the M2 recurrent couplings, and transitions between consecutive attractors are driven by low-dimensional noise fluctuations arising in the feedback projections from the subcortical nucleus back to M2. As a consequence of the stochastic origin of the transitions, the distribution of dwell times for each attractor is right-skewed, closely matching the empirical data. This model's prediction was confirmed in the data, where the ensemble fluctuations around each pattern were found to be low-dimensional and oriented in the direction of the next pattern in the sequence. This model presents a new interpretation for low-dimensional (differential) correlations: although their presence in sensory cortex may be detrimental for sensory encoding \citep{moreno2014information}, their presence in motor circuits seems to be essential for motor generation during naturalistic behavior \citep{recanatesi2022metastable}.


\subsection{Open issues}

The hypothesis that preparatory activity for upcoming actions is encoded in discrete attractors in M2 has been convincingly demonstrated using causal manipulations in head-fixed preparations in mice \lm{(see Appendix \ref{box:attractors})}, although
a causal test of this hypothesis in freely moving animals is currently lacking. A shortcoming of the metastable attractor model of action timing in \citep{recanatesi2022metastable} is the unidentified subcortical structure where the low-dimensional variability is originating. Thalamus and basal ganglia are both likely candidates as part of a large reciprocally connected mesoscopic circuit underlying action selection and execution (see Fig. \ref{figfive}B) and more work is needed to precisely identify the origin of the low-dimensional variability.

\lm{The metastable attractor model assumes the existence of discrete units of behavior at the level of actions, although large variability in movement execution  originating downstream of cortical areas may blur the distinction between the discrete behavioral units (see Discussion for more details). The extent to which behavior can be interpreted as a sequence of discrete behavioral units or, rather, a superposition of continuously varying poses (see Appendix \ref{box:videos} for an in-depth discussion of this issue) is currently open for debate. }

At higher levels of the behavioral hierarchy, repetitions of the same behavioral sequence (such as a jump attempt, an olfactory search trial, or a waiting trial, Fig. \ref{figtwo} and \ref{figthree}) exhibit large temporal variability as well, characterized by right-skewed distributions \citep{lottem2018activation}. \lm{It remains to be examined whether temporal variability in sequence duration may originate from a hierarchical model where sequences themselves are encoded in slow-switching metastable attractors in a higher cortical area or a distributed mesoscopic circuit (see Fig. \ref{figfive}C and Fig. \ref{figeight}).}

\section{Contextual modulation of temporal variability}

The second source of temporal variability in naturalistic behavior arises from contextual modulations, which can be {\it internally or externally driven}. When internally driven, they may arise from changes in brain or behavioral state such as arousal, expectation, or task engagement. When externally driven, they may arise from changes in environmental variables, from the experimenter's imposed task conditions, or from manipulations such as pharmacological, optogenetic or genetic ones. Contextual modulations can affect several qualitatively different aspects of behavioral units at each level of the hierarchy (actions, sequences, activities): \lm{Average duration; Usage frequency; Transition probabilities between units.} Moreover, context may also change the motor execution of a behavioral unit, for example by improving the vigor of a certain movement upon learning or motivation. For each type of modulation we will give several examples and review computational mechanisms that may explain them.

\subsection{Action timing}

The distribution of self-initiated action durations typically exhibits large variability, whose characteristic timescale can be extracted from their average duration (Fig. \ref{figthree}). The average timing of an action is strongly modulated by contextual factors, both internally and externally driven. Examples of internally generated contextual factors include expectation and history-effects. When events occur at predictable instants, anticipation improves performance such as reaction times. This classic effect of expectation was documented in an auditory two-alternative choice task \citep{jaramillo2011auditory}, where freely moving rats were rewarded for correctly discriminating the carrier frequency of a frequency-modulated target sound immersed in pure-tone distractors (Fig. \ref{figfour}A). The target could occur early or late within each sound presentation and temporal expectations on target timing were modulated by changing the ratio of trials with early or late targets within each block. When manipulating the expectations about sound timing, valid expectations accelerated reaction times and improved detection accuracy, showing enhanced perception. The auditory cortex is necessary to perform this task and firing rates in auditory cortex populations are modulated by temporal expectations.

Contextual changes in brain state may also be induced by varying levels of neuromodulators. In the self-initiated waiting task of Fig. \ref{figthree}A, optogenetic activation of serotonergic neurons in the Dorsal Raphe nucleus selectively prolonged the waiting period, leading to a more patient and less impulsive behavior, but did not affect the timing of other self-initiated actions \citep{fonseca2015activation}. 
Internally generated contextual modulations include history effects, which can affect the timing of self-initiated actions. In an operant conditioning task, where freely moving mice learned to press a lever for a minimum duration to earn a reward \citep{fan2012mechanisms}, the distribution of action timing showed dependence on the outcome of the preceding trial. After a rewarded trial, mice exhibited longer latency to initiate the next trial, but shorter press durations; after a failure, the opposite behavior occurred, with shorter latency to engage and longer press durations. Trial history effects are complex and action-specific and depend on several other factors, including prior movements, and wane with increasing inter-trial intervals. Inactivation experiments showed that these effect rely on frontal areas such as medial prefrontal and secondary motor cortices \citep{murakami2014neural,murakami2017distinct,schreiner2021mice}. 

Although the contextual modulations considered so far occur on a fast timescale of a few trials, they may also be the consequence of associative learning. In a lever press task, mice learned to adjust the average duration of the lever press to different criteria, in three different conditions where lever presses were always rewarded regardless of duration, or only rewarded if longer than 800ms or 1600ms. Within each of the three conditions, the distribution of action timing exhibited large temporal variability, yet the average duration was starkly different between the three criteria, as the mice learned the different criteria  \citep{fan2012mechanisms,schreiner2021mice}.  \lm{Reaction times \lm{to sensory stimuli} and self-initiated waiting behavior, in the form of long lever presses or nose-pokes, has emerged as a fruitful approach to test hypotheses on contextual modulations and decision-making in naturalistic scenarios.}

\subsection{Controlling action timing via neuronal gain modulation}

\begin{figure}[t]
\includegraphics[width=\linewidth]{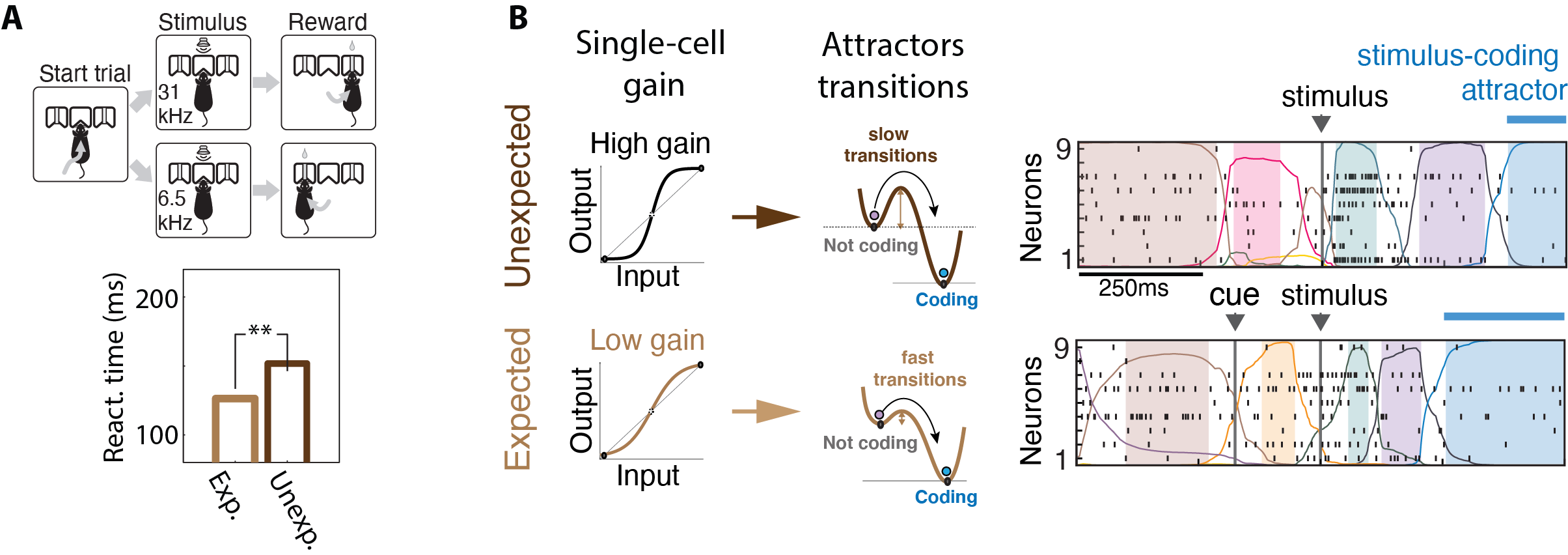}
\caption{{\it Contextual modulations mediated by changes in internal states.} {\bf A)} Expectation modulates reaction times. Top: Freely moving rats initiated were trained to choose either side port depending on the frequency features of a presented stimulus (a target frequency embedded in a train of distractors) to collect reward. Bottom: Reaction time for targets that were expected (light brown) were faster than unexpected (dark brown). {\bf B)} Expectation induces faster stimulus-coding. Contextual modulations that accelerate stimulus coding and reaction times may operate via a decrease in single-cell intrinsic gain (left), lowering the energy barrier separating non-coding attractors to the stimulus-coding attractor (center). Lower barriers allow for faster transitions into the stimulus coding attractor, mediating faster encoding of sensory stimuli in the expected condition compared to the unexpected condition. Right: Representative ensemble activity from rats gustatory cortex in two trials where the taste delivery was expected (bottom) or unexpected (top). The onset of the taste-coding attractor (blue) occurs earlier when the taste delivery is expected. Panel A adapted from Fig. 1 and 2  \citep{jaramillo2011auditory}, with permission from Elsevier. Panel B partially reproduced from Fig. 5 of \citep{mazzucato2019expectation}, with permission from Nature Publishing Group. They are not covered by the CC-BY 4.0 license and further reproduction of this panel would need permission from the copyright holder.
}
\label{figfour}
\end{figure}

The paramount role of contextual modulations in regulating action timing during naturalistic scenarios has been well documented. However, the neural mechanisms underlying these effects remain elusive. Results from head-fixed preparations revealed some possible explanations, which have the potential to generalize to the freely moving case. Within the paradigm of metastable attractors (see Appendix \ref{box:metaattractors}), the speed at which cortical activity encodes incoming stimuli can be flexibly controlled in a state-dependent manner by transiently changing the baseline level of afferent input currents to a local cortical circuit. These baseline changes may be driven by top-down projections from higher cortical areas or by neuromodulators. In a recurrent circuit exhibiting attractor dynamics, changes in baseline levels modulate the average transition times between metastable attractors \citep{mazzucato2019expectation,wyrick2021state}. In these models, attractors are represented by potential wells in the network's energy landscape, and the height of the barrier separating two nearby wells determines the probability of transition between the two corresponding attractors (lower barriers are easier to cross and lead to faster transitions, Fig. \ref{figfour}B). \lm{Changes in input baseline that decrease (increase) the barrier height lead to faster (slower) transitions to the coding attractor, in turn modulating reaction times.}

Although it is not possible to measure potential wells directly in the brain, using mean field theory one can show that the height of these potential wells is directly proportional to the neuronal gain as measured by single-cell transfer functions (for an explanation see Appendix \ref{box:metaattractors}). In particular, a decrease (increase) in pyramidal cells gain can lead to faster (slower) average action timing. This biologically plausible computational mechanism was proposed to explain the acceleration of sensory coding observed when gustatory stimuli are expected, compared to when they are delivered as a surprise \citep{samuelsen2012effects,mazzucato2019expectation}; \lm{and the faster encoding of visual stimuli observed in V1 populations during locomotion periods compared to when the mouse sits still \citep{wyrick2021state}. }

How can a neural circuit learn to flexibly adjust its \lm{responses to stimuli or the timing of self-initiated actions}? Theoretical work has established gain modulation as a general mechanism to flexibly control network activity in recurrent network models of motor cortex \citep{stroud2018motor}. Individual modulation of each neuron's gain can allow a recurrent network to learn a variety of target outputs through reward-based training and to combine learned modulatory gain patterns to generate new movements. After learning, cortical circuits can control the speed of an intended movement through gain modulation and affect the shape or the speed of a movement. Although the model in \citep{stroud2018motor} could not account for the across-repetition temporal variability in action timing, it is tempting to speculate that a generalization of this learning framework to incorporate the metastable attractor models of \citep{mazzucato2019expectation} could allow a recurrent circuit to learn flexible gain-modulation via biologically plausible synaptic plasticity mechanisms \citep{litwin2014formation}. This hypothetical model could potentially explain the contextual effects of learning on action timing observed in \citep{fan2012mechanisms,schreiner2021mice}, and the acceleration of reaction times in presence of auditory expectations (Fig. \ref{figfour}A \citep{jaramillo2011auditory}). Although this class of models has not been directly tested in freely moving assays, we believe that pursuing this promising direction could lead to important insights.

\subsection{Behavioral sequences}

\begin{figure}[t]
\begin{fullwidth}
\includegraphics[width=0.95\linewidth]{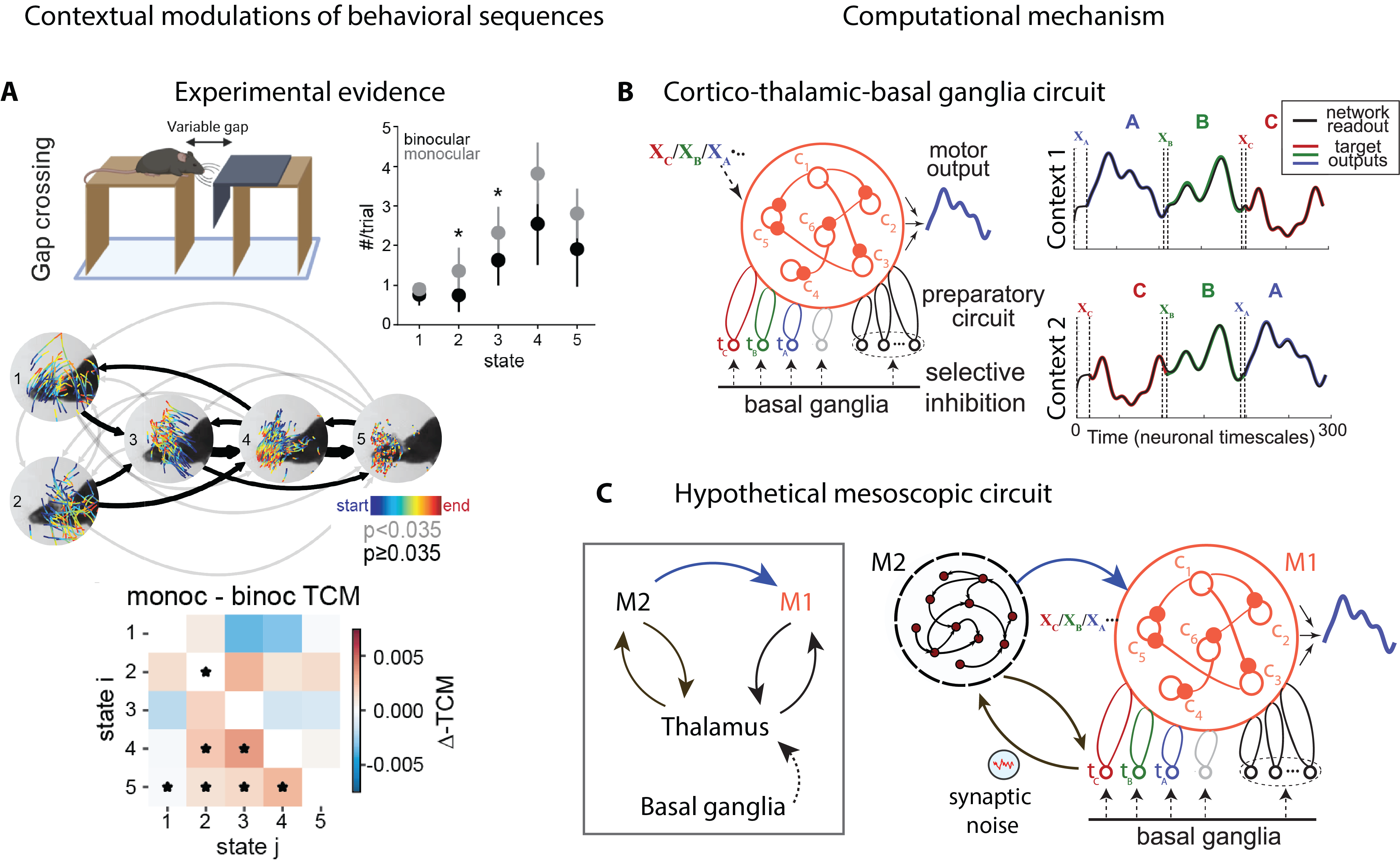}
\caption{{\it Contextual modulations of behavioral sequences.} {\bf A)} Freely moving mice were rewarded for successfully jumping across a variable gap. Center: Example traces of eye position from five movement states labeled with AR-HMM fit to pose tracking points during the decision period (progressing blue to red in time \lm{within the same state; states 2-3 represent vertical head movements}). Top right: Frequency of \lm{occurrence of} each state for binocular (black) and monocular (gray) conditions. Bottom: Difference between monocular and binocular transition count matrices; red transitions are more frequent in monocular, blue in binocular. {\bf \lm{B})} Left: Cortex-thalamus-basal ganglia circuit for behavioral sequence generation. The basal ganglia projections select thalamic units ($t_{A,B,C}$) needed for either motif execution or preparation. During preparation, the cortical population $c_i$ also receives an input $x_m$ specific to the upcoming motif m. Right: Generation of sequences of arbitrary orders, using preparatory periods (between vertical dashed lines) before executing each motif. {\bf \lm{C})} Hypothetical cortex-thalamus-basal ganglia circuit for behavioral sequence generation combining the metastable attractor model (Fig. \ref{figtwo}C-D) with the model of panel C. Secondary motor cortex (M2) provides the input to primary motor cortex (M1), setting the initial conditions for each motif. Synaptic noise in the feedback loop from thalamus to M2 generates temporal variability in action timing. Panels B and C are adapted from Fig. 4 of \citep{logiaco2021thalamic}. They are not covered by the CC-BY 4.0 license and further reproduction of this panel would need permission from the copyright holder. Panel A reproduced from Fig. 3 of  \citep{parker2021distance}.
}
\label{figfive}
\end{fullwidth}
\end{figure}

Contextual modulations of temporal variability may affect other aspects of natural behavior beyond average action timing, such as the frequency of occurrence of an action ("state usage") or the transition probabilities between actions within a behavioral sequence (Fig. \ref{figfive}A). These effects can be uncovered by analyzing freely moving behavioral videos with state space models, such as the auto-regressive hidden Markov model \citep{wiltschko2015mapping} ( \lm{Fig. \ref{figtwo}B, Fig. \ref{figfive}A and Appendix \ref{box:videos}}). Differences in state usage and transition probabilities between conditions may shed light onto the computational strategies animals deploy to solve complex ethological tasks, such as those involving sensorimotor integration. For example, in a distance estimation task where mice were trained to jump across a variable gap, \lm{a comparison of monocular and binocular mice revealed the different visually-guided strategies mice may use to perform a successful jump.} Mice performed more vertical head movements under monocular conditions compared to control (Fig. \ref{figfive}A, \lm{states 2 and 3 occur more frequently in the monocular condition}), revealing a reliance on motion parallax cues \citep{parker2021distance} .

During ongoing periods, in the absence of a task, animal behavior features a large variety of actions and behavioral sequences (Fig. \ref{figtwo}). Experimentally controlled manipulations can lead to strong changes in ongoing behavior reflected both in changes of state usage and of transition probabilities between actions, resulting in different repertoires of behavioral sequences. Examples of manipulations include: exposing mice to innately aversive odors and other changes in their surrounding environment; optogenetic activation of corticostriatal pathways \citep{wiltschko2015mapping}; and pharmacological treatment \citep{wiltschko2020revealing}. In the latter study, a classification analysis predicted with high accuracy which drug and specific dose was administered to the mice from a large panel of compounds at multiple doses. Comparison of \lm{state usage and transition rates} can also reveal subtle phenotypical changes in the structure of ongoing behavior in genetically modified mice compared to wild type ones. This phenotypic fingerprinting has led to insights into the behavior of mouse models of autism spectrum disorder \citep{wiltschko2015mapping,wiltschko2020revealing,klibaite2021deep}.

\subsection{Computational mechanisms underlying flexible behavioral sequences}

Recent studies have begun to shed light on the rules that may control how animals learn and execute behavioral sequences. These studies revealed various type of contextual modulations such as changes in the occurrence of single actions or in the transition probabilities between pairs of actions and proposed potential mechanisms underlying these effects. Biologically plausible models of mesoscopic neural circuits can generate complex sequential activity \citep{logiaco2021thalamic,murray2017learning}. In a recent model of sequence generation \citep{logiaco2021thalamic}, an extensive library of behavioral motifs and their flexible rearrangement into arbitrary sequences relied on the interaction between motor cortex, basal ganglia, and thalamus. In this model (Fig. \ref{figfive}B), the basal ganglia sequentially disinhibit motif-specific thalamic units, which in turn trigger motif preparation and execution via a thalamocortical loop with the primary motor cortex (M1). Afferent inputs to M1 set the initial conditions for motif execution. This model represents a biologically plausible neural implementation of a switching linear dynamical system \citep{linderman2017bayesian,nassar2018tree}, a class of generative models whose statistical structure can capture the spatiotemporal variability in naturalistic behavior (see Appendix \ref{box:videos}. 

How do animals learn context dependent behavioral sequences? Within the framework of corticostriatal circuits, sequential activity patterns can be learned in an all-inhibitory circuit representing the striatum \citep{murray2017learning}. Learning in this model is based on biologically plausible synaptic plasticity rules, consistent with the decoupling of learning and execution suggested by lesion studies showing that cortical circuits are necessary for learning, but that subcortical circuits are sufficient for expression of learned behaviors \citep{kawai2015motor}. This model can achieve contextual control over temporal rescaling of the sequence speed and facilitate flexible expression of distinct sequences via selective activation and concatenation of specific subsequences. Subsequent work uncovered a new computational mechanism underlying how motor cortex, thalamus and the striatum coordinate their activity and plasticity to learn complex behavior on multiple timescales \citep{murray2020remembrance}. The combination of fast cortical learning and slow subcortical learning may give rise to a covert learning process through which control of behavior is gradually transferred from cortical to subcortical circuits, while protecting learned behaviors that are practiced repeatedly against overwriting by future learning.

\subsection{Open issues}

The computational mechanisms discussed so far can separately account for some specific features of contextual modulations, but none of the existing models can account for all of them. Metastable attractor models explain \lm{how the large variability in action timing may be implemented from correlated noise (Fig. \ref{figthree}) and how contextual modulations of action timing may arise from neuronal gain modulation (Fig. \ref{figfour}C). However, it is not known whether such models can explain the flexible rearrangement of actions within a behavioral sequence (Fig. \ref{figfive}A-C)}. Conversely, models of flexible sequence execution and learning \citep{logiaco2021thalamic,murray2017learning,stroud2018motor} can explain the latter effect, but do not incorporate temporal variability in action timing across repetition. We would like to propose a hypothetic circuit model that combines  the cortex-basal ganglia-thalamic model of Fig. \ref{figfive}B together with the metastable attractor model of preparatory activity of Fig. \ref{figtwo}C-D to provide a tentative unified model of temporally variable yet flexible behavioral sequences. In this hypothetical mesoscopic circuit for flexible movement preparation and execution (Fig. \ref{figfive}C), the secondary motor cortex (M2), whose metastable attractors encode upcoming actions, provide the afferent inputs to M1 which set the initial conditions for movement execution. Additional feedback loops between M2 and the thalamus (already present in the model of Fig. \ref{figthree}D) may coordinate the timing of transitions across the whole circuit. Presynaptic noise in the thalamus-to-M2 projections implement the variability in action timing via noise-driven transitions between M2 attractors.  Although this model has not been studied yet, it can provide a natural and parsimonious explanation for the contextual modulations of behavioral sequences (Fig. \ref{figfive}A-B) while at the same time capturing the variability in action timing via gain modulation in M2 attractors (Fig. \ref{figtwo}). \lm{Moreover, it represents a direct circuit implementation of the state space models of behavior based on Markovian dynamics: discrete states representing actions/motifs correspond to discrete M2 attractors, whereas the continuous latent trajectories underlying movement execution correspond to low-dimensional trajectories of M1 populations.} This class of models can serve as a useful testing ground for generating mechanistic hypotheses and guide future experimental design. 

It remains to clarify the exact extent to which the \lm{stochastic} and contextual variability are independent sources. In principle, one could hypothesize that detailed knowledge and control of all experimental variables and behavioral state and their history might explain part of the trial-to-trial variability as contextual variability conditioned on these variables. For example, in the Marshmallow task about 10\% of the temporal variability in waiting times could be attributed to a trial-history effect, which relied on an intact medial prefrontal cortex and was abolished with its inactivation \citep{murakami2017distinct}. Moreover, in a subset of sessions this fraction of variability could be predicted by the activity of transient neurons before trial onset \citep{murakami2014neural,murakami2017distinct}. The remaining 90\% of the unexplained across-repetition variability was attributed to a stochastic mechanism, likely originating from metastable dynamics \citep{recanatesi2022metastable}. In general, it is an open question to investigate whether what we think of as noise driving trial-to-trial variability could just be another name for a contextual variable that we have not yet quantified. Alternatively, \lm{stochastic} variability might be genuinely different from contextual variability and originate from noise inherent in neural spiking or other activity-dependent mechanisms.
%
%

\section{Hierarchical structure}

The temporal organization of naturalistic behavior exhibits a striking hierarchical structure \citep{tinbergen2020study,dawkins1976hierarchical,simon1991architecture}, where actions are nested into behavioral sequences which are then grouped into activities. Higher levels in the hierarchy emerge at longer timescales: actions/movements occur on a subsecond scale, behavioral sequences span at most a few seconds and activities last for longer periods of several seconds to minutes. The crucial aspect of this behavioral hierarchy is its complexity: an animal's behavior unfolds along all timescales simultaneously. What is the organization of this vast spatiotemporal hierarchy? What are the neural mechanisms supporting and generating this \lm{nested} temporal structure?

\subsection{\lm{A case study: {\it C. elegans} locomotor behavior}}

\begin{wrapfigure}{l}{.5\textwidth}
\includegraphics[width=\hsize]{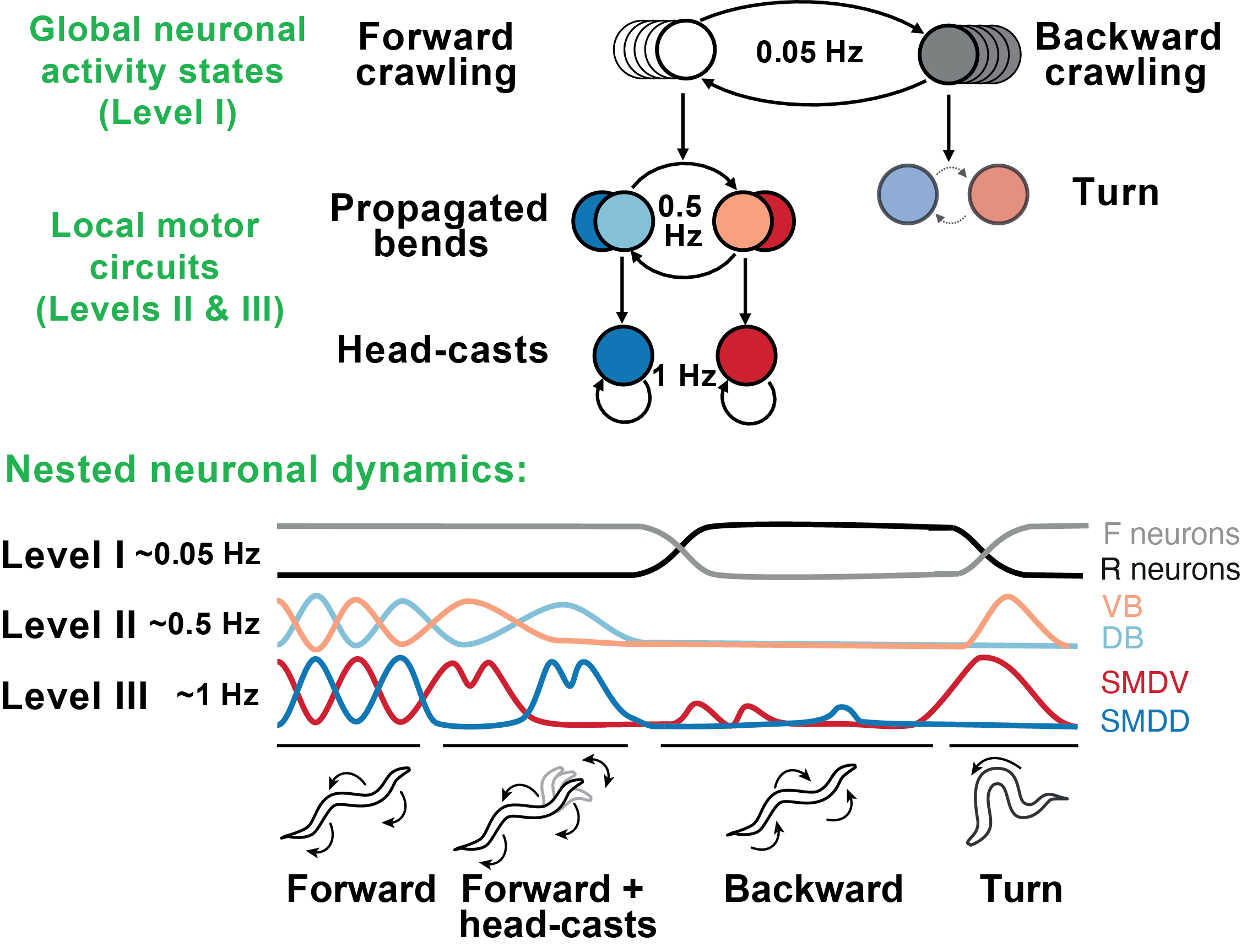}
\caption{\lm{{\it Behavioral and neural hierarchies in C. elegans}. Top: Behavioral hierarchy in C. elegans. A 0.05-Hz cycle drives switches between forward and reverse crawling states, with intermediate level 0.5-Hz crawling undulations, and lower level 1-Hz head-casts. Bottom: Slow dynamics across whole-brain circuits reflect upper-hierarchy motor activity; fast dynamics in motor circuits drive lower-hierarchy movements. Slower dynamics tightly constrain the state and function of faster ones. Adapted from Fig. 8 \citep{kaplan2020nested}, with permission from Elsevier. They are not covered by the CC-BY 4.0 license and further reproduction of this panel would need permission from the copyright holder.}}
\label{fig:worm}
\end{wrapfigure}

\lm{Remarkable progress in both behavioral and neurophysiological aspects of the hierarchy has been made in the worm {\it C. elegans} (Fig. \ref{fig:worm}). The worm locomotor behavior exhibits a clear hierarchical organization with three distinct timescales: sub-second body-bends and head-casts, second-long crawling undulations, and slow reverse-forward cycles \citep{gomez2016hierarchical}. At the neuronal level, this hierarchy is generated by nested neuronal dynamics where upper-level motor programs are supported by slow activity spread across many neurons, while lower-level behaviors are represented by fast local dynamics in small multi-functional populations \citep{kaplan2020nested}. Persistent activity driving higher-level behaviors gates faster activity driving lower-level behaviors, such that, at lower levels, neurons show dynamics spanning multiple timescales simultaneously \citep{hallinen2021decoding}. Specific lower-level behaviors may only be accessed via switches at upper levels, generating a non-overlapping, tree-like hierarchy, in which no lower-level state is connected to multiple upper-level states.}

\lm{At the top of this motor hierarchy we find a much longer-lasting organization of states in terms of exploration, exploitation, and quiescence. In contrast to the strict, tree-like structure observed in the motor hierarchy, lower level motor feature are shared across these states, albeit with different frequency of occurrence. Whereas the motor hierarchy is directly generated by neuronal activity, this state-level hierarchy may rely on neuromodulation \citep{ben2009molecular,flavell2013serotonin}}

\subsection{Behavioral hierarchies in flies and rodents}

\lm{How much of this tight correspondence between behavioral and neural hierarchies generalizes from worms to insects and mammals?} We will start by considering the first two levels of the hierarchy, namely, how actions concatenate into sequences. Are action sequences organized as a chain? Alternatively, is there a hierarchical structure where individual actions, intermediate subsequences, and overall sequences can be flexibly combined? A chain-like organization would require a single controller  concatenating actions, but could be prone to error or disruption. A hierarchical structure could be error tolerant and flexible, at the cost of requiring controllers at different levels of the hierarchy. A recent experimental {\it tour de force} demonstrated the existence of a hierarchical structure in the learning and execution of heterogeneous sequences in a self-paced operant task \citep{geddes2018optogenetic}. Mice were trained to perform a "penguin dance" consisting of a sequence of two or three consecutive left lever presses (LL or LLL) followed by a sequence of two or more right lever presses (RR or RRR) to obtain a reward. Mice acquired the sequence hierarchically rather than sequentially, a learning scheme that is inconsistent with the classic reinforcement learning paradigm, which predicts sequence learning occurs in the reverse order of execution \citep{sutton2018reinforcement}. Using closed-loop optogenetic stimulation, the authors revealed the differential roles played by striatal direct and indirect pathways in controlling, respectively, the expression of a single action (either L or R), or a fast switch from one subsequence to the next (from the LL block to RR, and from RR to the reward approach).

\begin{figure}[t]
\begin{fullwidth}
\includegraphics[width=0.95\linewidth]{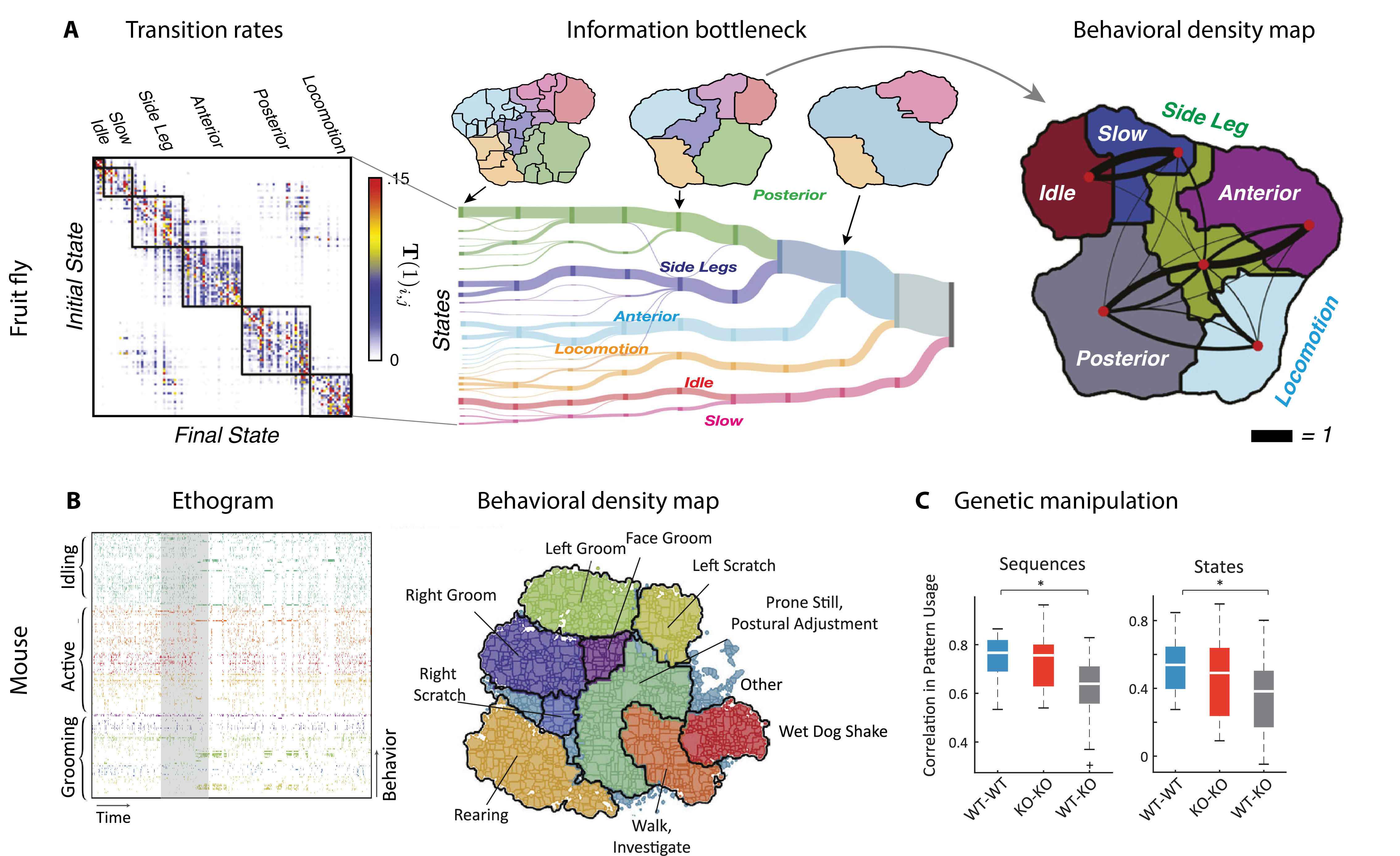}
\caption{{\it Hierarchical structure in freely moving fly and mouse behavior.} {\bf A)} \lm{Hierarchial variability in the fruit fly behavior. The Markov transition probability matrix (left) between postures reveals a clustered structure upon applying the predictive information bottleneck with six clusters (black outline on the left plot). Center: Hierarchical organization for optimal solutions of the information bottleneck  for predicting behavior on increasingly slower timescales (varying clusters from 25 to 1, left to right; colored vertical bars are proportional to the percentage of time a fly spends in each cluster). Right: Behavioral clusters are contiguous in behavioral space (same clusters as in the transition matrix in the left panel).} {\bf B)} Left: Rat behavior Is hierarchically organized into behavioral sequences and states. A temporal pattern matching algorithm detected repeated behavioral patterns in freely moving mice behavioral recordings. Right: \lm{Data from 16 rats co-embedded in a two dimensional t-SNE behavioral map was clustered with a watershed transform, revealing behavioral categories segregated to different regions of the map. The ethogram on the left is annotated from these behavioral clusters.} {\bf C)} Sequence and state usage probabilities for wild-type and Fmr1-KO rats show a significantly decreased correlation between different genotypes. Panel A adapted from Fig. 2 and 5 \citep{wiltschko2015mapping}, with permission from PNAS. Panels B and C are adapted from Fig. 3, 4 and 6 of \citep{marshall2021continuous}, with permission from Elsevier. They are not covered by the CC-BY 4.0 license any: further reproduction of this panel would need permission from the copyright holder.}
\label{figsix}
\end{fullwidth}
\end{figure}

Is this hierarchical structure confined to short behavioral sequences, or is it an organizing feature of behavior at all timescales? By analyzing videos of ground-based fruit fly during long sessions of spontaneous behavior (Fig. \ref{figtwo}B), the authors of \citep{berman2014mapping,berman2016predictability} were able to classify the behavioral space of freely moving flies, identifying a hundred stereotyped, frequently reoccurring actions, interspersed with bouts of non-stereotyped behaviors. Recurring behavioral categories emerged as peaks in the behavioral space probability landscape, labelled as walking, running, head grooming, wing grooming (Fig. \ref{figsix}A). In order to uncover the organization of behavior at different timescales, the authors estimated which behavioral representations (movements, sequences, or activities?) could optimally predict the fly's future behavior on different temporal horizons (from 50ms to minutes), applying the information bottleneck method \citep{tishby2000information}. This predictive algorithm revealed multiple time scales in the fly behavior, organized into a hierarchical structure reflecting the underlying behavioral programs and internal states. The near future could be optimally predicted from segmenting behavior according to actions at the fastest level of the hierarchy. The optimal representation of behavior which could optimally predict the distant future up to minutes away was based on slower, coarser groups of actions grouped into activities. These longer timescales manifest as nested blocks in the transition probability matrix, implying that even though behavior looks Markovian at short timescales \lm{(i.e., it is efficiently captured by a transition probability matrix between different movements)}, a strongly non-Markovian structure emerges on longer timescales \citep{alba2020exploring} (Fig. \ref{figsix}A). 

A hallmark of the fly behavior emerging from this analysis is that the different branches of this hierarchical tree in the fruit fly are segregated. Namely, movements/actions occurring during grooming do not occur during locomotion, and so on for all different activities. This multiscale representation of the fly behavior was leveraged to dissect the descending motor pathways in the fly. Optogenetic activation of single neurons during spontaneous behavior revealed that most of the descending neurons drove stereotyped behaviors which where shared by multiple neurons and often depended on the behavioral state prior to activation \citep{cande2018optogenetic}. An alternative statistical approach based on a hierarchical hidden Markov model revealed that although all flies use the same set of low-level locomotor features, individual flies vary considerably in the high-level temporal structure of locomotion, and how this usage is modulated by different odors \citep{tao2019statistical}. \lm{This behavioral idiosyncrasy of individual-to-individual phenotypic variability has been traced back to specific genes \citep{ayroles2014behavioral} regulating neural activity \citep{buchanan2015neuronal} in the central complex of the fly brain.}

This multiscale analysis of ongoing behavior was applied to mice in a recent study, where mouse body piercing allowed for tracking of the three-dimensional pose with high accuracy (Fig. \ref{figsix}B \citep{marshall2021continuous}). Examining the behavioral transition matrix at different timescales, signature of non-Markovian dynamics peaked at 10 to 100 second timescales. On 15 seconds timescales, pattern sequences emerged featuring sequentially ordered actions, such as grooming sequences of the face followed by the body. On a minute-long timescales, states of varying arousal or task engagement emerged lacking stereotyped sequential ordering. \lm{Applying a watershed transform to the density map revealed the emergence of a similar hierarchy as in the fly, with segregated behavioral categories at long timescales. A notable difference between flies and rodents was that,} while fly behavior was very consistent across different animals, mouse hierarchies featured strong signatures of individuality in behavioral kinematics, usage, and patterning. \lm{This behavioral idiosyncrasy of individual mice was similar to the complex movement sequences, akin to short "dances," which mice learned in order to produce prescribed long lever-pressing in an interval timing task \citep{kawai2015motor}.} While previous studies of contextual modulations of freely moving rodents by pharmacological or genetic maniputations were mostly confined to comparison of action usage statistics \citep{wiltschko2015mapping}, this unprecedented access to the whole behavioral hierarchy allowed a multiscale assessment of contextual effects. In a rat model of the fragile X syndrome, while mutant and wild-type rats had similar locomotor behavior, the former showed abnormally long grooming epochs, characterized by different behavioral sequences and states compared to wild types (Fig. \lm{\ref{figsix}}C). This encouraging pilot study highlights the advantages of multiscale comparative taxonomy of naturalistic behavior to investigate behavioral manifestations of complex conditions such as autism spectrum disorder.

\subsection{Multiple timescales of neural activity}

\begin{figure}[t]
\begin{fullwidth}
\includegraphics[width=\linewidth]{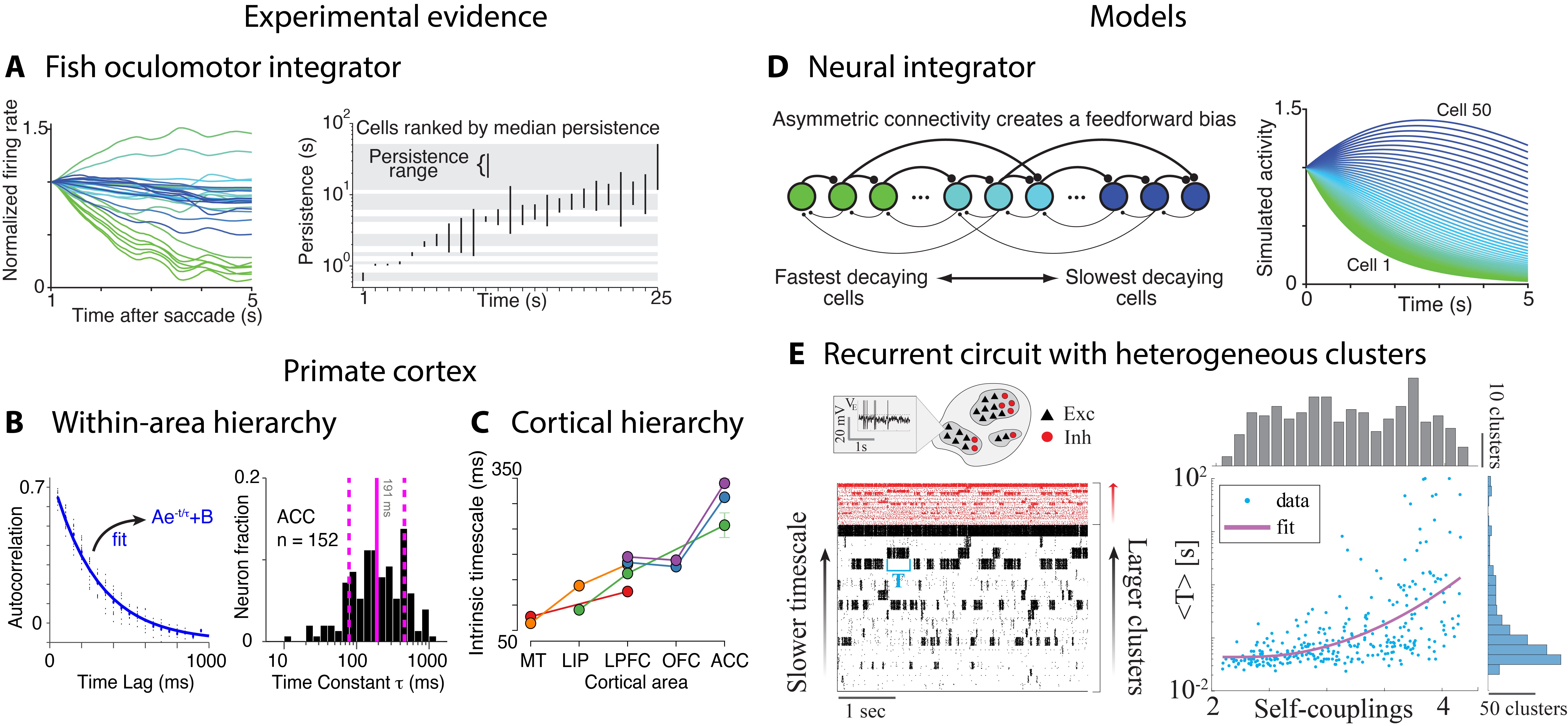}
\caption{{\it Computational principles underlying \lm{the heterogeneity of timescales}.} {\bf A)} Firing rate of neurons imaged
in zebrafish larvae (left, colored according to the rostrocaudal and dorsoventral neuron location) reveal the correspondence between persistence time (right, each bar represent the persistence time range for each cell) and location along these dimensions. {\bf B)} Left: Autocorrelation of an example neuron from orbitofrontal cortex (OFC) in awake monkey (blue: exponential fit). Right: Histogram of the time constants reveals a large variability across OFC neurons (solid and dashed vertical lines represent mean and SD). {\bf C)} Intrinsic timescales across the visual-prefrontal hierarchy in five data sets estimated as the average population autocorrelation. {\bf D)} Persistent activity with heterogeneous timescales in the oculomotor system can be explained by a progressive filtering of activity propagating down a circuit including a mixture of feedback and  functionally feedforward interactions, realizing a uniformly detuned line attractor. \lm{{\bf E)} A heterogeneous distribution of timescales naturally emerges in a recurrent network of excitatory (black) and inhibitory (red) spiking neurons arranged in clusters, generating time-varying activity unfolding via sequences of metastable attractors (left: representative trial, clusters activate and deactivate at random times; neurons are sorted according to cluster membership). Larger clusters (at the top) activate for longer intervals.  Right: The distribution of cluster activation timescales $<T>$, proportional to a cluster size, exhibits a large range from 20ms to 100s.}  Panel A adapted from Fig. 4 and 8 \citep{miri2011spatial}, with permission from Nature Publishing Group. Panel C and D are adapted from Fig. 1 of \citep{murray2014hierarchy} and Fig. 8 of \citep{miri2011spatial}, with permission from Nature Publishing Group. They are not covered by the CC-BY 4.0 license and further reproduction of this panel would need permission from the copyright holder. Panel B adapted from Fig. 2 of  \citep{cavanagh2016autocorrelation}. Panel E adapted from Fig. 5 of  \citep{stern2021reservoir}.
}
\label{figseven}
\end{fullwidth}
\end{figure}

What are the neural mechanisms generating the hierarchy of timescales observed during naturalistic behavior? Is there evidence that neural activity is simultaneously varying over multiple timescales? Although no studies directly addressed these questions yet, a number of experimental and theoretical approaches have provided evidence for multiple timescales of neural activity. Some evidence for temporal heterogeneity in neural activity was reported in restrained animals during stereotyped behavioral assays. A heterogeneous distribution of timescales of neural activity was found in the brainstem of the restrained larval zebrafish, by measuring the decay time constant of persistent firing across a population of neurons comprising the oculomotor velocity-to-position neural integrator (Fig. \ref{figseven}A, \citep{miri2011spatial}). The decay times varied over a vast range 0.5s-50s across cells in individual larvae. This heterogeneous distribution of timescales was later confirmed in the primate oculomotor brainstem \citep{joshua2013diversity}, suggesting that timescale heterogeneity is a common feature of oculomotor integrators conserved across phylogeny. Single neuron activity may also encode a long memory of task-related variables. In head-fixed monkeys performing a competitive game task, temporal traces of the reward delivered in previous trials were encoded in single neuron spiking activity in frontal areas over a wide range of timescales, obeying a power law distribution up to 10 consecutive trials  \citep{bernacchia2011reservoir}. 

The timescale of intrinsic fluctuations in spiking activity can be also estimated from single-neuron spike autocorrelation functions (Fig. \ref{figseven}B). In awake head-fixed primates, during periods of ongoing activity the distribution of intrinsic autocorrelation timescales within the same cortical circuit was found to be right-skewed and approximately lognormal (Fig. \ref{figseven}B), in the range from 10ms to 1 second \citep{cavanagh2016autocorrelation}. Moreover, comparison of the population-averaged autocorrelations during ongoing periods revealed a hierarchical structure across cortical areas, varying from 50ms to 350ms along the occipital-to-frontal axis (\citep{murray2014hierarchy} Fig. \ref{figseven}C).

While all these results were obtained in restrained animals, it is an open question whether neural activity during naturalistic behavior exhibit temporal hierarchies similar to those observed in behavior. Current evidence from freely moving rodents engaged in waiting tasks \citep{murakami2017distinct,schreiner2021mice,recanatesi2022metastable} revealed the presence of multiple timescales in single-neuron activity from secondary motor and prefrontal cortices. These timescales range from the subsecond scale (single attractors), to a few seconds seconds (attractor sequences), to tens of seconds or minutes (trial-history dependence). It is tantalizing to speculate that even longer timescales may be present for neural activity to be able to generate the vast hierarchy of timescales observed in naturalistic behavior (Fig. \ref{figsix}). Evidence from associative learning tasks in rodents found that population activity in hippocampal CA1 and CA3 encodes multiplexed information about several aspects of the task occurring on multiple trials such as context, place, value, and objects \citep{mckenzie2014hippocampal}. Although an explicit analysis of temporal correlations was not carried out in this study, these results suggest that a hierarchy of timescales may be present in the hippocampus and emerge during learning.

\subsection{Computational mechanisms underlying heterogeneous distributions of timescales}

What are the computational mechanisms underlying this hierarchical temporal structure featuring a wide distribution of timescales? Single-cell biophysical properties such as differences in membrane time constant across cell types or synaptic time constants across receptors (e.g., faster AMPA and GABA vs. slower NMDA receptors) may generate multiple timescales in sub-second range, from a few milliseconds to a few hundred milliseconds \citep{gjorgjieva2016computational}. Although this class of single-cell mechanisms might explain the temporal hierarchy observed in the posterior-anterior cortical axis \citep{murray2014hierarchy} by relying on systematic differences in cell type specific features, they are unlikely to explain the much wider temporal hierarchy \lm{observed across neurons within the same area (Fig. \ref{figseven}A-B), or the even larger hierarchy observed in behavior (Fig. \ref{figsix}).} 

Theoretical studies highlighted the central role played by recurrent synaptic couplings within a local circuit for generating long timescales emerging from the recurrent dynamics. Recurrent networks with random recurrent couplings can generate time-varying neural activity whose timescale may be tuned to very slow values at a critical point \citep{toyoizumi2011beyond,magnasco2009self}. Long timescales may also emerge in random neural networks with an excess of symmetric couplings \citep{marti2018correlations}.  \lm{In these examples, all cells share the same timescale. Although this common timescale can be tuned to arbitrarily long values, these circuits are incapable of giving rise to temporal heterogeneity either at the single cell or at the population level.} 

Two biologically plausible ingredients were shown to be sufficient to generate long timescales with temporal heterogeneity in their distributions: recurrent couplings realizing local functional neural clusters and heterogeneity in synaptic couplings. The first requirement of functional assemblies is a connectivity motif ubiquitously observed in biological circuits, namely the fact that the strength of recurrent couplings decays with spatial distance between pairs of neurons, leading to the emergence of local functional assemblies of strongly coupled neurons \citep{Song2005-mj,Perin2011-ts,Lee2016-ld,kiani2015natural,miller2014visual}. In recurrent linear network models, functional assemblies can generate neural activity exhibiting slow relaxation times following stimulation. In order to generate a heterogeneity in the distribution of relaxation times, functional assemblies can be coupled by heterogeneous long-range connections, arranged along a spatial feedforward gradient (Fig. \ref{figseven}D) \citep{miri2011spatial,joshua2013diversity,chaudhuri_xjw2014diversity}. Linear networks featuring both local functional assemblies and spatial heterogeneity in long-range couplings were able to reproduce the heterogeneous decay times found in the brainstem oculomotor integrator circuit in zebrafish and primates \citep{miri2011spatial,joshua2013diversity} and in the reward integration times in the macaque cortex \citep{bernacchia2011reservoir}. It remains an open question whether these integrator models could be generalized to explain hierarchical activity in motor generation, perhaps stacking multiple layers of them. A shortcoming of these models is the fact that they require fine tuning of synaptic couplings and do not generate attractor dynamics, which recent experimental evidence suggest is the basic building block of preparatory motor activity \citep{inagaki2019discrete,recanatesi2022metastable}. Other network models generating multiple timescales of activity fluctuations  were  proposed based on self-tuned criticality with anti-hebbian plasticity \citep{magnasco2009self}, or multiple block-structured connectivity \citep{merav2015blocks}.

An alternative robust and biologically plausible way to generate a vast hierarchy of timescales was proposed based on the ingredients of recurrent functional clusters and heterogeneity in synaptic couplings (Fig. \ref{figseven}E) \citep{stern2021reservoir}, two common features observed across cortical circuits \citep{marshel2019cortical,kiani2015natural,miller2014visual,Song2005-mj,Perin2011-ts,Lee2016-ld}, supported by theoretical evidence \citep{LitwinKumarDoiron2012,wyrick2021state}. \lm{In this model, excitatory and inhibitory neurons are arranged in clusters of heterogeneous sizes, generating metastable activity whose typical timescale is measured by the on-off cluster switching time (see Appendix \ref{box:metaattractors}). In this model, a cluster's timescale is proportional to its size and larger clusters exhibit longer timescales, yielding a heterogeneous distribution of timescales in the range observed in cortex (Fig. \ref{figseven}A-C).} An appealing feature of this model is that it could be generalized to other domains which exhibit fluctuations simultaneously varying over a large range of timescales, such as spin glasses \citep{bouchaud1992weak}, metabolic networks of E. coli \citep{almaas_barbasi2004metabolic_flux_ecoli,emmerling2002metabolic} and yeast cultures \citep{roussel2007yeast_chaos_observation,aon2008yeast_scalefree}. It remains an open question whether the relationship between a neural cluster's size and its intrinsic timescale is realized in cortical circuits and whether it can explain the origin of the hierarchical variability in naturalistic behavior.

\subsection{Future directions: Neural mechanisms generating temporal hierarchies}

Future studies should address how the hierarchy of timescales found in behavior may emerge from computational mechanisms. There are several missing links along this path. First, although temporally heterogeneous neural activity was found in several brain areas and species under different experimental conditions, no study to date investigated the presence of temporal hierarchies in neural activity during naturalistic behavior. All the necessary tools are available: recent advances in neurotechnology demonstrate the feasibility of chronic recordings of large neural populations during freely moving behavior \citep{juavinett2019chronically,van2021implantation}; simultaneously performing behavioral classification analyses from pose tracking software \citep{mathis2018deeplabcut,marshall2021continuous,pereira2019fast}. This first piece of the puzzle is thus within reach.

\begin{figure}[t]
\includegraphics[width=0.95\linewidth]{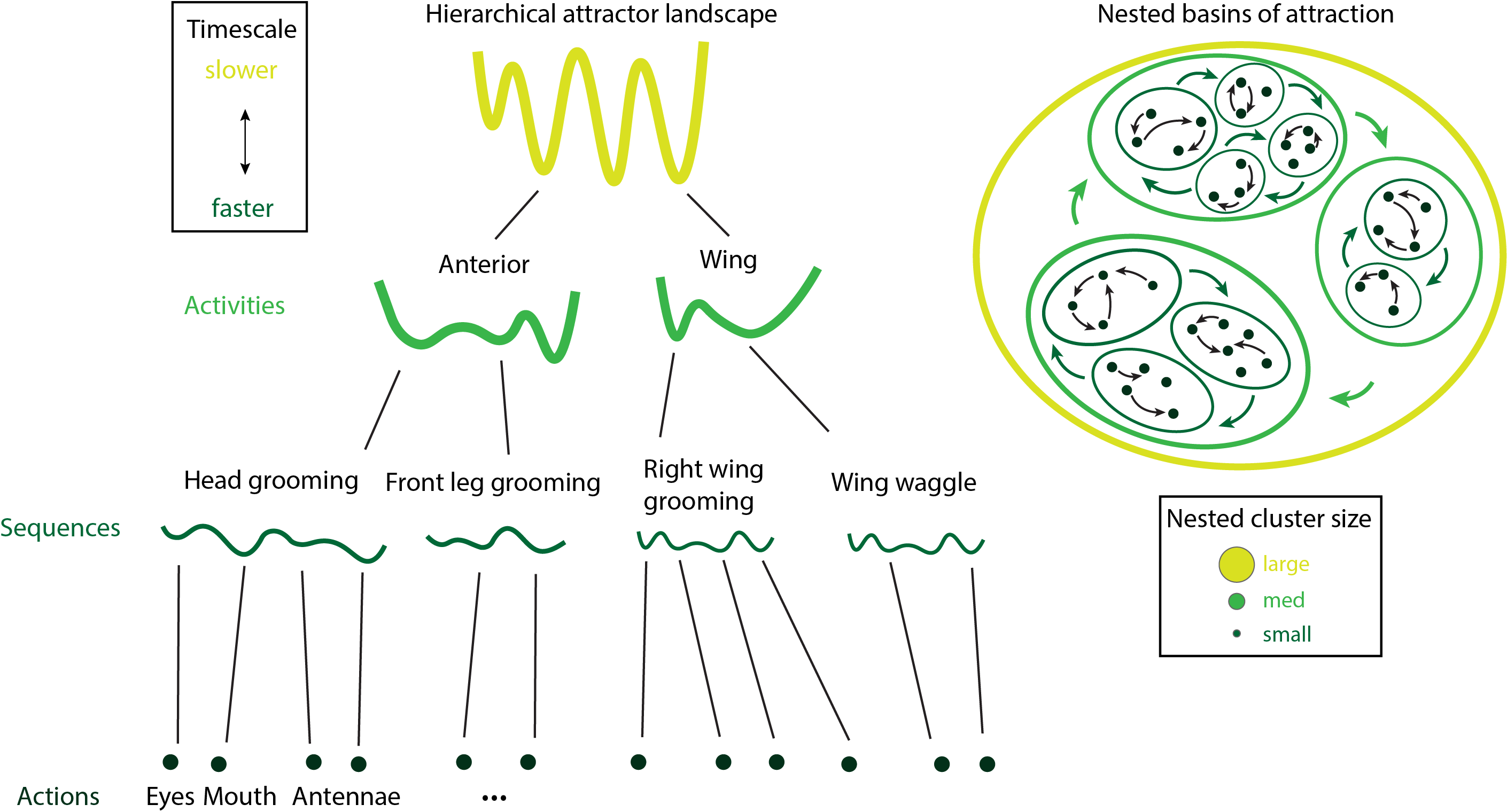}
\caption{{\it Hierarchical attractor networks can explain \lm{a tree-like behavioral hierarchy.}} A computational framework to explain the hierarchical structure of behavior discovered in the fruit fly and in rodents (Fig. \ref{figsix}). Attractors encoding for actions, sequences and activities (left) are supported by nested basins of attractions (right), where the size of a basin determines the intrinsic timescale of the corresponding activity. Actions have small basins, corresponding to fast timescales, while activities have large basins, corresponding to slow timescales. Attractors are non-overlapping, consistent with the tree-like structure of the behavioral hierarchy.}
\label{figeight}
\end{figure}

On the theory side, although recurrent dynamics can generate heterogeneous distributions of timescales (Fig. \ref{figseven}E), this model needs to be extended to a fully realistic framework for explaining \lm{nested} temporal hierarchies in naturalistic behavior such as the proposed mesoscopic circuit in Fig \ref{figfive}C. Here, we propose a simple roadmap to bridge this gap, building on some of the theoretical ideas we reviewed above. This theory is based on a hierarchical structure that, for lack of a better term, we will denote {\it "Attractors all the way down"} (Fig. \ref{figeight}). We start from the observation that self-initiated actions within a behavioral sequence are represented as metastable attractors in secondary motor cortex \citep{inagaki2019discrete,recanatesi2022metastable} (Fig. \ref{figthree}). We then extend this observation to a general principle positing that behavioral units at each level of the temporal hierarchy (not only actions, but also sequences, activities) are represented by metastable attractors. The stochasticity in behavioral unit duration can be achieved by generating transitions between attractors at each level via low-dimensional variability, arising from mesoscopic feedback loops involving cortex and subcortical areas \citep{recanatesi2022metastable} (Fig. \ref{figthree}D). The average transition time between behavioral units at each level will then depend on the barrier height separating the corresponding attractors \citep{cao2016collective,mazzucato2019expectation,wyrick2021state}. As one moves up the hierarchy from actions (fast transitions = lower barriers) to sequences and activities (slow transitions = higher barriers) the potential wells separating the corresponding attractors become deeper and the basins of attraction wider. \lm{This increase in timescales can be achieved in a biologically manner by assuming that increasingly larger neural populations encode for slower behavioral features \citep{stern2021reservoir} (see Fig. \ref{figseven}E), which can be realized in a biologically plausible way as a hierarchy of clusters within clusters \citep{schaub2015emergence}.}  Contextual modulations in average duration of a behavioral unit at each level of the hierarchy can be implemented by top-down changes of the barrier heights between attractors via gain modulation (Fig \ref{figfour}C). The new theoretical ingredient required for this theory to work is the presence of a nested structure in the attractor landscape, whereas the basin of attraction of lower level attractors (actions) are contained within the basins of higher level attractors (sequences) all the way up to the largest basins representing long-term activities. \lm{The architecture of this model provides some immediate predictions for neural activity. First, neurons exhibit conjunctive selectivity to multiple variables at different levels of the hierarchy, which has been observed for example in rodent hippocampus during freely moving tasks \citep{mckenzie2014hippocampal} and in primate prefrontal cortex during complex decision-making tasks \citep{rigotti2013importance}. Second, single cell activity is modulated by many attractors, in agreement with the multi-stable activity observed in sensory \citep{mazzucato2015dynamics} and frontal areas \citep{recanatesi2022metastable} where neuronal representations of attractors are dense rather than sparse. Third, fluctuations in the activity of single neurons should vary over multiple timescales {\it simultaneously}, encoding information about multiple levels of the hierarchy; this is the structure that was observed in {\it C. elegans} where neurons representing movements exhibit both fast and for slow fluctuations correlated to multiple levels of the behavioral hierarchy \citep{kaplan2020nested} (Fig. \ref{fig:worm}).}

Which network architectures can implement a hierarchical attractor landscape? The behavioral hierarchy discovered in \lm{flies \citep{berman2016predictability} and rodents \citep{marshall2021continuous}} can be approximated as a tree: behavioral units at a lower level of the hierarchy only occur during a specific unit at a higher level of the hierarchy  (Fig. \ref{figsix}). For example, movements occurring during the fruit fly locomotion activities never occur during idle activities. This structure is typical of phylogenetic classification in taxonomy, where the distance from the root to the leaves is the same for every leaf and the tree is called ultrametric \citep{rammal1986ultrametricity}. It is not known whether  behavioral trees are ultrametric; in order to clarify this structure, a notion of distance in behavioral space will have to be introduced and examined. A classic model of hierarchical attractors, which could potentially capture the tree-based hierarchy, is an extension of the Hopfield network where the stored pattern are hierarchically arranged into a tree \citep{parga1986ultrametric} (Fig. \ref{figeight}). This class of models were originally proposed to explain the effect of word semantic categorization during memory encoding, storage and retrieval. \lm{A contamination between the domains of naturalistic behavior and natural language processing has been proposed early on \citep{dawkins1976hierarchical} and recently applied to describe {\it C. elegans}  \citep{gomez2016hierarchical,gupta2019context} and larval zebrafish behavior \citep{reddy2022lexical}.}

What are potential neural implementations of this network architecture? In the case of mammals, at the lowest level of the hierarchy, neural populations in the rodent secondary motor cortex (M2) generate metastable attractors representing upcoming actions \citep{inagaki2019discrete,recanatesi2022metastable}. Other subcortical areas, such as the basal ganglia or the thalamus, are likely involved in the generation of complex behavioral sequences \citep{murray2017learning,logiaco2021thalamic,recanatesi2022metastable}. How are higher order behavioral units, such as sequences and activities, encoded? One possibility is that they could also be encoded in M2 attractors, as recent evidence suggests M2 populations encode for a wide range of behavioral variables \citep{cazettes2021reservoir}. We hypothesize an architecture featuring nested assemblies where larger populations, whose activity varies over progressively slower timescales \citep{stern2021reservoir,schaub2015emergence}, hierarchically encode for slower features of behavior (from actions to sequences to activities). Experimental evidence shows that single neurons in M2 are selective to multiple actions \citep{recanatesi2022metastable}, suggesting that mixed selectivity \citep{rigotti2013importance} could play an important role in generating the high-dimensional hierarchical attractor landscape necessary to capture the complexity of naturalistic behavior. An alternative architecture might involve a distributed cortical circuit where the neural representations of behavioral units at different levels of the hierarchy are encoded in multiple frontal cortical areas, such as action sequences in M2, and activities in prefrontal cortex, which is known to control behavior on longer timescales such as trial-history effects \citep{murakami2014neural,murakami2017distinct,schreiner2021mice}. In both scenarios, we hypothesize that while behavioral units are encoded as cortical attractors, transitions between attractor rely on feedback loops involving cortico-subcortical circuits \citep{murray2017learning,logiaco2021thalamic,recanatesi2022metastable}.

\section{Challenges and opportunities}
\lm{
In this review we mostly focused on behavioral variability generated by central planning circuits, which can be explained at least in part by metastable attractors in premotor areas, encoding for preparatory activity, as part of a larger mesoscopic circuits including thalamic nuclei and the basal ganglia. There are other sources of variability we have not addressed, such as that originating from movement execution, from stochastic individuality, and from long-term processes such as circadian rhythms or aging. In this Section, we will discuss these alternative sources, and then assess the shortcomings of the current theoretical framework, missing links in the current literature and potential avenues for future research.
\subsection{Other sources of behavioral variability}
{\it Variability in movements and body posture.} A large source of variability in behavior originates from variability in movement execution \citep{dhawale2017role}. In a delayed response task in overtrained primates, preparatory neural activity in premotor areas could only account for half of the trial-to-trial variability in movements, potentially ascribing the rest to variability in movement execution \citep{churchland2006central}. It is challenging to dissect this finer movement variability from the state-space model approach to behavioral classification, as it aims at capturing discrete stereotypical movement features. In order to capture the finer scale of movements and body posture in C. elegans  \citep{schwarz2015changes,szigeti2015searching}, powerful methods from dynamical systems such as state-space reconstruction have been deployed \citep{costa2019adaptive,costa2021maximally,ahamed2021capturing} (these are different from the "state-space models" from statistics). This variability may originate from the motor periphery, such as noise in force production within muscles \citep{van2004role}. \\
{\it Stochastic individuality.} One important source of behavioral variability is phenotypic variability across different individuals with identical genetic profile, an important aspect of behavior in ethological and evolutionary light \citep{honegger2018stochasticity}. Stochastic individuality is defined as the part of the phenotypic variability in non-heritable effects which cannot be predicted
from measurable variables such as learning or other developmental conditions - such as behavioral differences in identical twins reared in the same environment. Signatures of stochastic individuality seem prevalent in rodents \citep{kawai2015motor,marshall2021continuous}, much more so than in flies \citep{berman2014mapping}. Existing theoretical models have not examined which neural mechanisms may underlie this individuality. This variability may arise  from differences in developmental wiring of brain circuits related to axonal growth. The metastable attractor model in \citep{recanatesi2022metastable} may naturally accommodate for some stochastic individuality. The location of each attractor in firing rate space is drawn from a random distribution, so that across-animals variability may stem from different random realization of the attractor landscape with the same underlying hyperparameters (i.e., mean and covariance of the neural patterns).\\
{\it Variability on longer timescales.} Internal states, such as hunger \citep{corrales2016internal}, and circadian rhythms \citep{patke2020molecular}  induce daily modulations of several aspects of behavior. In the fruit fly, different clusters of clock neurons are implicated in regulating rhythmic behaviours, including, wake-sleep cycles, locomotion, feeding, mating, courtship, and metabolism. Activation of circadian clock neurons in different phases of the cycle drives expression of specific behavioral sequences and targeted manipulations of particular clusters of clock neurons is sufficient to recapitulate those sequences artificially \citep{dissel2014logic,yao2014drosophila}. In mammals, the circadian pacemaker is located in the hypothalamus, interacting with a complex network of neuronal peripheral signals downstream of it \citep{hastings2018generation,schibler2015clock}. Other sources of contextual variability include variations in the levels of hormones and neuromodulators, which modulate behavioral sequences on longer timescales \citep{nelson2005introduction}. Although hormones and neuromodulators may not directly drive expression of specific behaviors, they typically prime an animal to elicit hormone-specific responses to particular stimuli in the appropriate behavioral context, as observed prominently during social behavior such as aggression and mating \citep{schibler2015clock}. Although the behavioral effects of these long-term sources of contextual modulations have not been included in current models of neural circuits, the theoretical framework based on attractor networks in Fig. \ref{figfive}C could be augmented to account for them. An afferent higher order cortical area, recurrently connected to M2, may toggle between different behavioral sequences and control long-term variations in their expression, for example via gain modulation. A natural candidate for this controller area is the medial prefrontal cortex, which is necessary to express long-term biases in the waiting task (Fig. \ref{figthree}) \citep{murakami2017distinct}. In this augmented model, a top-down modulatory input to the higher order controller, representing afferent inputs from circadian clock neurons or neuromodulation, may modulate the expression of different behavioral sequences implementing these long-term modulations.\\
{\it Aging}. Other sources of contextual modulation of behavior may act on even longer timescales spanning the whole lifetime of an individual \citep{churgin2017longitudinal}, such as homeostasis and development. The interplay between these two mechanisms may explain how the motor output of some neural circuits maintains remarkable stability, in the face of the large variability in neural activity observed across a population, or in the same individual during development \citep{prinz2004similar,bucher2005animal}. Across the entire lifespan, different brain areas develop, mature and decline at different moments and to different degrees \citep{sowell2004mapping,yeatman2014lifespan}. Connectivity between areas develops at variable rates as well \citep{yeatman2014lifespan}. It would be extremely interesting to investigate whether these different mechanisms, unfolding over the lifespan of an individual, can be accounted for within the framework of multi-area attractor networks presented above.
\subsection{Social behavior and contextual modulations}
Although most of the experiments discussed in this review entailed the behavior of individual animals, contextual modulations of behavior were prominently observed in naturalistic assays comprising the interaction between pairs of animals, such as hunting and social behaviors. In the prey-capture paradigm, a mouse pursues, captures, and consumes live insect prey \citep{hoy2016vision,michaiel2020dynamics}. Prey-capture behavior was found to strongly depend on context. Experimental control of the surrounding environment revealed that mice rely on vision for efficient prey-capture. In the dark, the hunting behavior is severely impaired: only at close range to the insect is the mouse able to navigate via auditory cues. Another remarkable example of context dependent social behavior was demonstrated during male-female fruit fly mating behavior \citep{calhoun2019unsupervised}. During courtship bouts, male flies modulate their songs using specific feedback cues from their female partner such as their relative position and orientation. A simple way to model the relationship between sensory cues and the choice of a specific song in terms of linear 'filters' \citep{coen2014dynamic}, where a common assumption is that the sensorimotor map is fixed. Relaxing this assumption and allowing for moment-to-moment transition between more than one sensorimotor map (a GLM-HMM), the authors of  \citep{calhoun2019unsupervised} uncovered latent states underlying the mating behavior corresponding to different sensorimotor strategies, each strategy featuring a specific mapping from feedback cues to song modes. Combining this insight with optogenetic manipulation of specific neurons revealed that neurons previously thought to be command neurons for song production are instead responsible for controlling the switch between different internal states, thus regulating the courtship strategies. Finally, a tenet of naturalistic behavior is vocal communication, which combines aspects of sensory processing and motor generation in the realm of complex social interactions. A particularly exciting model system is the marmoset, where new  techniques  to  record neural   activity in freely-moving animals during social behavior and vocalization \citep{nummela2017social} together with newly developed optogenetic tools \citep{macdougall2016optogenetic} and multi-animal pose tracking algorithms \citep{pereira2020sleap,lauer2021multi} hold the promise to push the field into entirely new domains \citep{eliades2017marmoset}.
\subsection{Normative theory of behavioral variability} 
What is the beneficial role of behavioral variability? A basic aspect of motor variability is avoiding predation or competition. Seminal work in songbirds has established that motor variability can be actively generated and controlled by the brain for the purpose of learning \citep{olveczky2005vocal,kao2005contributions}. Remarkably, variability in song production is not simply due to intrinsic noise in motor pathways but is introduced into motor cortex analog RA by a dedicated upstream area LMAN (analog to a premotor area), which is required for song learning. Theoretical modeling further revealed a potential mechanism to generate this motor variability, relying on topographically organized projections from LMAN to RA to amplify correlated neural fluctuations \citep{darshan2017canonical}. Strikingly, this mechanism provides universal predictions for the statistics of babbling shared by songbirds and human infants. In humans, trial-to-trial motor variability can be interpreted as a means to update control policies and motor output within a reinforcement learning paradigm \citep{wu2014temporal}. What are the specific benefits of temporal variability? It is tantalizing to speculate that variable timing may be an adaptive feature of motor behavior. Beyond predation avoidance, timing variability may allow animals to explore the temporal aspects of a given sequence of behavior independently of the choices of actions. Animals could learn proper timing of an action sequence by a search in timing space independent of action selection and vice-versa. Future work should explore the advantages of temporal variability in driving learning of precise timing.
\subsection{New theoretical directions}
{\it Behavioral energy landscape.} The structure of naturalistic behavior emerging on long timescales in the fruit fly is a hierarchical tree-like organization (Fig. \ref{figsix}A), which is consistent with an underlying neural circuit architecture based on hierarchical basins of attractions (Fig. \ref{figeight}). Can we derive a representation of behavior as a complex energy landscape directly from the behavioral data itself? A promising approach is given by the probability density map approach \citep{berman2016predictability}. One could define the log-probability of the density map as an energy potential, where the transition rates between behavioral features yield a probability flux along this potential landscape. It is tantalizing to speculate that using the combination of energy gradient and probability flux one could derive a data-driven nonlinear dynamical system describing behavior \citep{wang2015landscape}. Other approaches to infer a potential energy landscape directly from data have been successfully applied to spike trains \citep{genkin2020moving,duncker2019learning}, although in a regime where the energy landscape has only a few minima. Alternatively, in chemical kinetics or fluid flow, methods based on the transfer operator formalism have been applied to find effective free energy landscapes and metastable states from experimental data or simulations \citep{bowman2013introduction}, and have been recently applied to animal behavior \citep{costa2021maximally}. \\
{\it Neural mechanisms of flexible behavioral hierarchies.} The model of nested basins of attractions proposed in Fig. \ref{figeight} can explain a tree-like behavioral hierarchy, where behavioral units at lower levels (e.g., actions) are not shared by different units at a higher level (e.g., sequences), as observed in fruit flies \citep{berman2016predictability} and mice \citep{marshall2021continuous}. Although the tree-like hierarchical organization emerged from assays where individual animals were monitored in isolation, recent studies of social behavior seem to challenge this structure. A large variety of qualitatively new behaviors arise from social interactions, including fighting, mating, and others. While some of these behaviors involve specialized behavioral units, others involve simultaneous execution of multiple behavioral units leading to multi-tasking. For example, during courtship, a male fly can simultaneously approach a female fly (locomotion) and sing, two behaviors that would be mutually exclusive in the absence of a female fly. As a consequence, the simple tree-like hierarchical organization of behavior observed in isolated individuals (Fig. \ref{figseven}) might break down during social interactions and lead to a flexible organization where actions are shared between multiple sequences and activities.\\
Are the computational mechanisms generating the tree-like hierarchy sufficient to generate flexible hierarchies? Recent theoretical work highlighted the importance of a neural circuit architecture that segregates the motor preparation and execution in different areas. In the cortico-thalamic-basal ganglia model of Fig. \ref{figfive}B, flexible sequences can be generated by rearranging existing motifs as well as by learning new motifs without interfering with previous ones \citep{logiaco2021thalamic}. Moreover, flexible sequences can be learned using biologically plausible Hebbian plasticity in the striatum \citep{murray2017learning}. These theoretical models provide an exciting blueprint for further investigation into learning and expression of flexible behavioral hierarchies. Recent work on the social behavior of bats suggest that investigating the joint activity of multiple interacting brains may help shed further light on these questions \citep{zhang2019correlated,zhang2021unifying}. \\
{\it A theory of metastable dynamics in biologically plausible models.} A variety of neural circuit models have been proposed to generate metastable dynamics (see Appendix \ref{box:metaattractors}). However, a full quantitative understanding of the metastable regime is currently lacking. Such theory is within reach in the case of recurrent networks of continuous rate units. In circuits where metastable dynamics arises from low-dimensional correlated variability \citep{recanatesi2022metastable}, dynamic mean field methods could be deployed to predict the statistics of switch times from underlying biological parameters. In biologically plausible models based on spiking circuits, it is not known how to quantitatively predict switching times from underlying network parameters. Phenomenological birth-death processes fit to spiking network simulations can give a qualitative understanding of the on-off cluster dynamics \citep{huang2017once,shi2022cortical}, and it would be interesting to derive these models from first principles. Mean field methods for leaky-integrate-and-fire networks can give a qualitative prediction of the effects of external perturbations on metastable dynamics, explaining how changes in an animal's internal state can affect circuit dynamics \citep{mazzucato2019expectation,wyrick2021state}. These qualitative approaches should be extended to fully quantitative ones. A promising method, deployed in random neural networks, is based on the universal colored-noise approximation to the Fokker-Planck equation, where switch times between metastable states can be predicted from microscopic network parameters such as neural cluster size \citep{stern2021reservoir}. Finally, a crucial direction for future investigation is to improve the biological plausibility of metastable attractor models to incorporate different inhibitory cell-types. Progress along this line will open the way to quantitative experimental tests of the metastable attractor hypothesis using powerful optogenetic tools.
}

\section{Acknowledgments}

I would like to thank Zach Mainen, James Murray, Cris Niell, Matt Smear, Osama Ahmed, the participants of the Computational Neuroethology Workshop 2021 in Jackson, and the members of the Mazzucato and Murray labs for many discussions and suggestions. I would like to acknowledge the reviewers for providing critical feedback and suggestions. LM was supported by National Institute of Neurological Disorders and Stroke grant R01-NS118461 and by National Institute on Drug Abuse grant R01-DA055439 (CRCNS).

\appendix

\begin{appendixbox}
\section{From behavioral videos to behavioral features.}
\label{box:videos}
\lm{
Several video analysis tools are available to extract behavioral features from videos, ranging from standard computer vision algorithms to deep learning-based methods. These algorithms are fast and reliable and allow to track animal movements at different levels of spatiotemporal resolution, typically yielding three different kinds of time-series outputs. Deep learning algorithms with human annotated video frames yield a set of coordinates in ambient space representing tracked points on an animal's body \citep{mathis2018deeplabcut,segalin2020mouse,pereira2019fast}. Unsupervised algorithms yield a set of low dimensional variables obtained from dimensionally reducing the image pixel space via principal component analysis or other nonlinear compression methods \citep{wiltschko2015mapping,batty2019behavenet}. Spectral decomposition of the videos obtained from a time-frequency analysis yield probability density maps \citep{berman2014mapping,szigeti2015searching}. Translating these low-dimensional observables into a meaningful representation of behavior for subsequent analyses requires further assumptions on the nature of feature representations, leading to difficult choices which should be carefully evaluated case by case. \\
{\bf Discrete vs continuous features.} The first choice regards the definition of what constitutes a fundamental unit of behavior for a particular model system and whether this unit is discrete or continuous. Is a "slow locomotion" bout distinct from a "fast locomotion" bout? Are they just different manifestation within the variability range of the same "locomotion" motif? This choice is related to the timescale at which behavior is analyzed to test a particular hypothesis and is based on the level of perceived stereotypy present in the observed behavior \citep{brown2018ethology,berman2018measuring,pereira2020quantifying}. At the sub-second timescale, movements can be represented as continuous low-dimensional trajectories of body parts. A continuous representation of behavior assumes that behavioral time series can be represented as a superposition of continuous behavioral motifs or "eigenshapes" and has been fruitful in quantifying the behavior of the worm {\it C. elegans} and the {\it Drosophila} larva \citep{stephens2008dimensionality,szigeti2015searching} (although discrete representations have been applied to the same systems \citep{brown2013dictionary,vogelstein2014discovery,schwarz2015changes}). Continuous trajectories of movements can be interpreted and investigated as complex dynamical systems as well \citep{ahamed2021capturing}. In the zebrafish swimming, clear peaks in distribution of kinematic parameters suggest categorical distinctions \citep{marques2018structure}, although for some movements the kinematics vary more continuously, allowing the fish to perform a smooth range of swim types. The approach where units of behavior are discrete and can be separated in time leads to representations based on ethograms, namely, graphs where nodes are discrete actions and edges transition rates between them \citep{bateson1976growing}. Ethograms implicitly assume that variability across repetitions of the same action ("stereotypy") is smaller than the variability between different actions. In order to bridge the continuous nature of movements with its inherent variability to a discrete interpretation in terms of actions, two successful approaches have been used. When using state space models (Fig. \ref{figtwo}A and \ref{figfive}B), the fundamental units of behavior can be captured by stereotyped, low-dimensional, autoregressive trajectories or linear dynamical systems (LDS). The behavioral time series unfolds as a sequence of piecewise LDS trajectories, where each trajectory represents a discrete action, yet it accommodates for large variability in its continuous expression. This approach has been successfully applied both in rodents (with auto-regressive hidden Markov models \citep{wiltschko2015mapping}) and worms (with switching LDS \citep{linderman2019hierarchical}). Two of its advantages are the fact that it provides a generative model which can be used to simulate synthetic behavior; and the fully Bayesian implementation of inference. One limitation is that the number of behavioral units must be specified as a hyperparameter, raising the thorny issue of model selection. In freely moving rodents, up to a dozen states were sufficient to capture animal behavior when performing a task \citep{findley2021sniff,parker2021distance}, but many more states (up to a hundred) were found during spontaneous behavior \citep{wiltschko2015mapping} (Fig. \ref{figtwo}A). Sample size  and the bias-variance tradeoff are important issues to consider in this class of models. 
An alternative data-driven approach to segmentation of behavior avoids having to choose the number of units and their dynamical features {\it a priori}, but rather lets the structure emerge from the data at different timescales. Using density-based clustering to obtain a distribution of time points in feature space, once can identify peaks in the density map via watershed transforms (Fig. \ref{figtwo}B). This approach has been successfully used in the adult {\it Drosophila} \citep{berman2014mapping} and in rodents \citep{marshall2021continuous}. The advantage of the watershed transform is that the segmentation into discrete units depends on the level of coarse-graining or the timescale at which the behavior is analyzed, allowing greater flexibility compared to state space models. Both approaches combine continuous representations of movements with discrete segmentation into behavioral units.\\
{\bf Non-Markovianity and long timescales}. Discrete behavioral representations based on state space models are convenient and effective for capturing naturalistic behavior with a limited number of actions and within a limited timescale horizon. By construction, their temporal structure is constrained to reveal only two timescale: the duration of a single movement or trajectory (on the sub-second scale) and the duration of a behavioral sequence encoded in the transition probability matrix (up to a few seconds). Moreover, they are based on the strong assumption that behavioral sequences are Markovian: the next action only depends on the previous one. A crucial limitation of this approach is thus their inability to extract spatiotemporal features on longer timescales. Because of this shortcoming, they fail to model the ubiquitous hierarchical structure observed in naturalistic behavior \citep{tinbergen2020study,dawkins1976hierarchical,simon1991architecture} (although, hierarchical state space models have been deployed \citep{tao2019statistical}). In order to account for the non-stationarity and non-Markovianity observed over long timescale, one would be forced to introduce hundreds of hidden states, challenging both the interpretability and inference ability of the model. If the focus of the investigation is to analyze longer timescale and non-stationarity/non-Markovianity in behavior, the unsupervised density-based clustering approach is a more principled way to proceed \citep{berman2016predictability,alba2020exploring}.
}
\end{appendixbox}

\begin{appendixbox}
\label{box:attractors}
\section{Attractors dynamics in neural circuits}

Attractors are a concept in the theory of dynamical systems. An attractor is defined as the set of states a system evolves to starting from a large  set of initial conditions, defined as its basin of attraction. The main feature of an attractor is that if the system is perturbed slightly, it tends to return to it. In the context of recurrent neural networks (RNN, Fig. A), attractors are configurations of the network activity that exhibit these features and they can be either time-independent (point attractors, Fig. A), time-varying but periodic (limit cycles), or time-varying but non-periodic (chaotic attractors). The classic example of an attractor neural network exhibiting point attractors is the Hopfield network \citep{hopfield1982neural,amit1985spin}, which was introduced as a model of auto-associative memory. The Hopfield network represents an RNN with $N$ units $x_i$ (for $i=1,\ldots,N$) evolving according to the dynamics $\tau\dot x_i=-x_i+\sum_{j=1}^N w_{ij}\phi(x_j)$, where $\tau$ and $\phi(x)$ are single-cell time constant and transfer function, the latter typically chosen with a saturating non-linearity in neuroscience models \citep{hopfield1984neurons}. In a Hopfield network, the synaptic couplings $w_{ij}=\sum_{\mu=1}^pM_i^\mu M_j^\mu$ encode the network attractors $\{M_i^\mu\}_{\mu=1}^p$.  After initializing the network activity at a random value $x^{0}_i$ (empty circle in Fig. A), the network temporal dynamics quickly converge to the attractor $M^\mu$ which is closest to $x^{0}$ (black circle in Fig. A). The set of initial conditions that converge to a particular attractor defines its basin of attraction, represented as potential wells in an energy landscape (in Fig. \ref{figattractor}A, the basin of the attractor of $M^2$ is represented by the values of $x$ such that $a<x<b$ ). 
If we interpret this model as an auto-associative memory circuit, then each attractor $M^\mu$ represents a stored memory; when presenting the network with a hint ($x^0$) that is close to the memory $M^2$, the network will quickly retrieve the full memory. A crucial feature of attractor dynamics is that when we slightly perturb the network activity away from the attractor (blue lightning in Fig. A), then if the new configuration (empty blue circle) still lies within the basin of attraction of $M^2$, the network will quickly converge back to the attractor. 

It is possible to generalize these Hopfield-like stable attractors to biologically plausible networks of spiking neurons with cell-type specific connectivity \citep{amit1997model}. These models explain the features of neural activity observed in several areas of the association cortex (inferotemporal cortex: \citep{fuster1981inferotemporal,miyashita1988neuronal}; prefrontal cortex: \citep{funahashi1989mnemonic,wilson1993dissociation}; posterior parietal cortex \citep{koch1989unit}). In these experiments featuring a delay-response task, a stimulus is presented and then withdrawn, followed by a delay period at the end of which the monkey performs a stimulus-specific saccade to obtain a reward. Neural activity in association areas exhibits stimulus-selective persistent activity during the delay period following the removal of the stimulus, which can be interpreted as originating from attractors in the underlying neural circuit. Such self-sustained persistent activity preserves an active memory of a visual stimulus after it is removed. Long-lasting persistent activity was also reported in the fruit fly following optogenetic stimulation and connectome reconstruction suggest this activity may be supported by strong recurrent couplings between persistently active neurons \citep{deutsch2020neural}. Recent advances with optogenetics in rodents further confirmed that delay persistent activity is stable against perturbations, a crucial feature of attractor dynamics. In a delay-response task (\citep{inagaki2019discrete,finkelstein2021attractor} Fig. B), neural activity recorded in the antero-lateral motor cortex (a part of the secondary motor cortex) showed persistent choice-selective activity during the delay period (dashed lines in Fig. B). In trials where bilateral optogenetic perturbations transiently silenced the local circuit at the beginning of the delay epoch (full lines in Fig. B), stimulus-selective population activity quickly recovered to the unperturbed values (full lines in Fig. B), suggesting that delay activity in this area can be interpreted as originating from attractors.

\begin{center}
\includegraphics[width=0.8\linewidth]{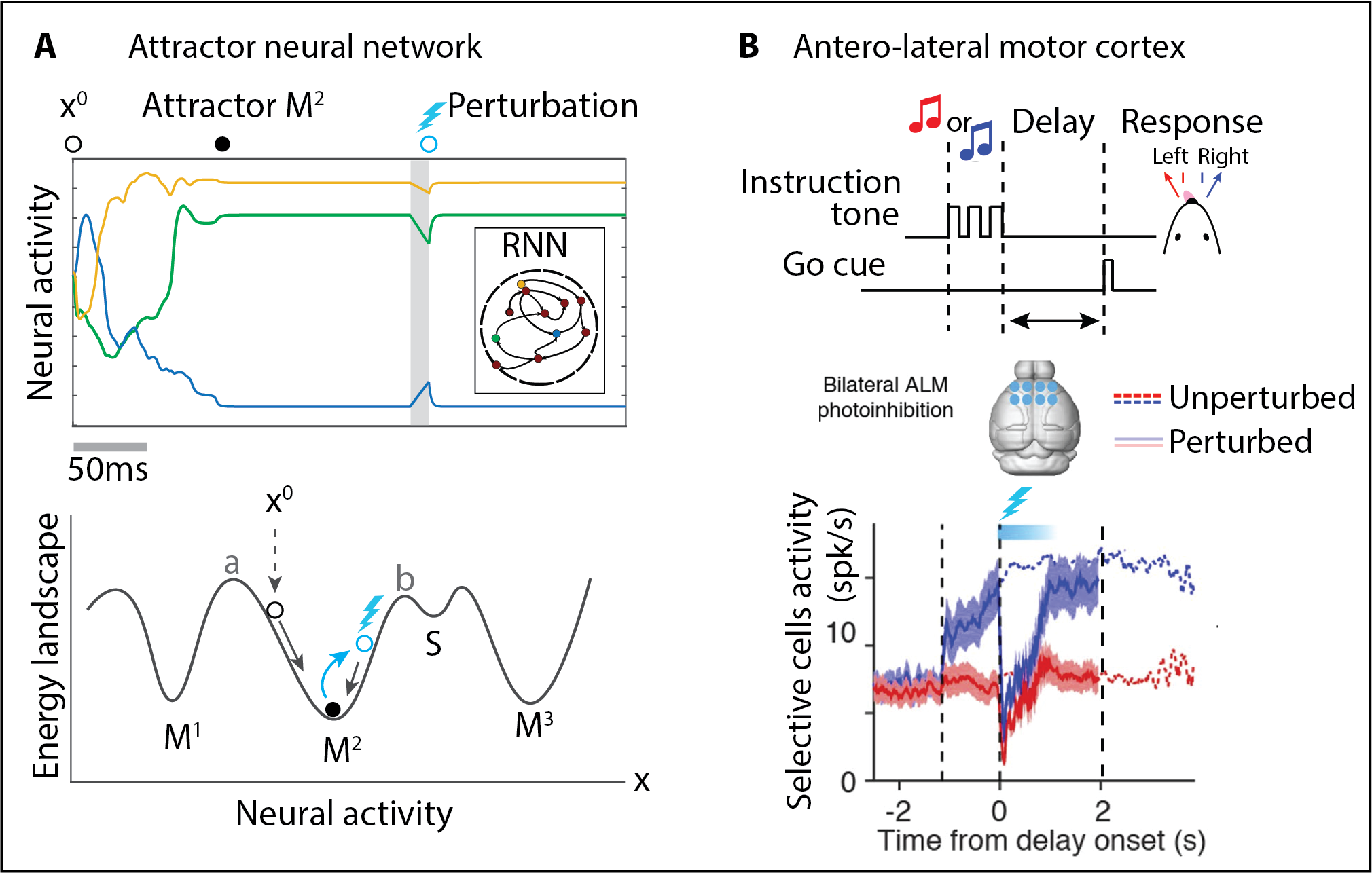}
\label{figattractor}
\captionof{figure}{{\bf A)} In an attractor neural network, the activity of three representative units starting from initial values $x^0$ rapidly converges to the closest attractor ($M^2$). A small perturbation displaces the activity away from the attractor (blue circle) and the dynamics quickly returns to the attractor. Bottom: Attractors are represented as potential wells in an energy landscape. Initial conditions or perturbations within the basin of attraction of the attractor $M^2$ (defined by $a<x<b$) quickly converge back to the attractor. {\bf B)} 
In a head-fixed delay response task (top), one of two tones was presented during the sample epoch and the mouse reported its decision after the delay epoch by directional licking (randomized delay duration). The anterolateral motor cortex was photoinhibited bilaterally during the first 0.6s of the delay epoch (cyan bar) and quickly returned to its unperturbed level. Mean spike rate of lick-right preferring neurons (blue and red thick lines: unperturbed lick-right and -left trials; dashed lines: perturbed trials). Panels C and D adapted from Fig. 1, 6 and S6 \citep{inagaki2019discrete}, with permission from Nature Publishing Group. They are not covered by the CC-BY 4.0 license and further reproduction of this panel would need permission from the copyright holder. 
}
\end{center}

\end{appendixbox}

\begin{appendixbox}
\section{Metastable attractors}
\label{box:metaattractors}

\begin{center}
\includegraphics[width=\linewidth]{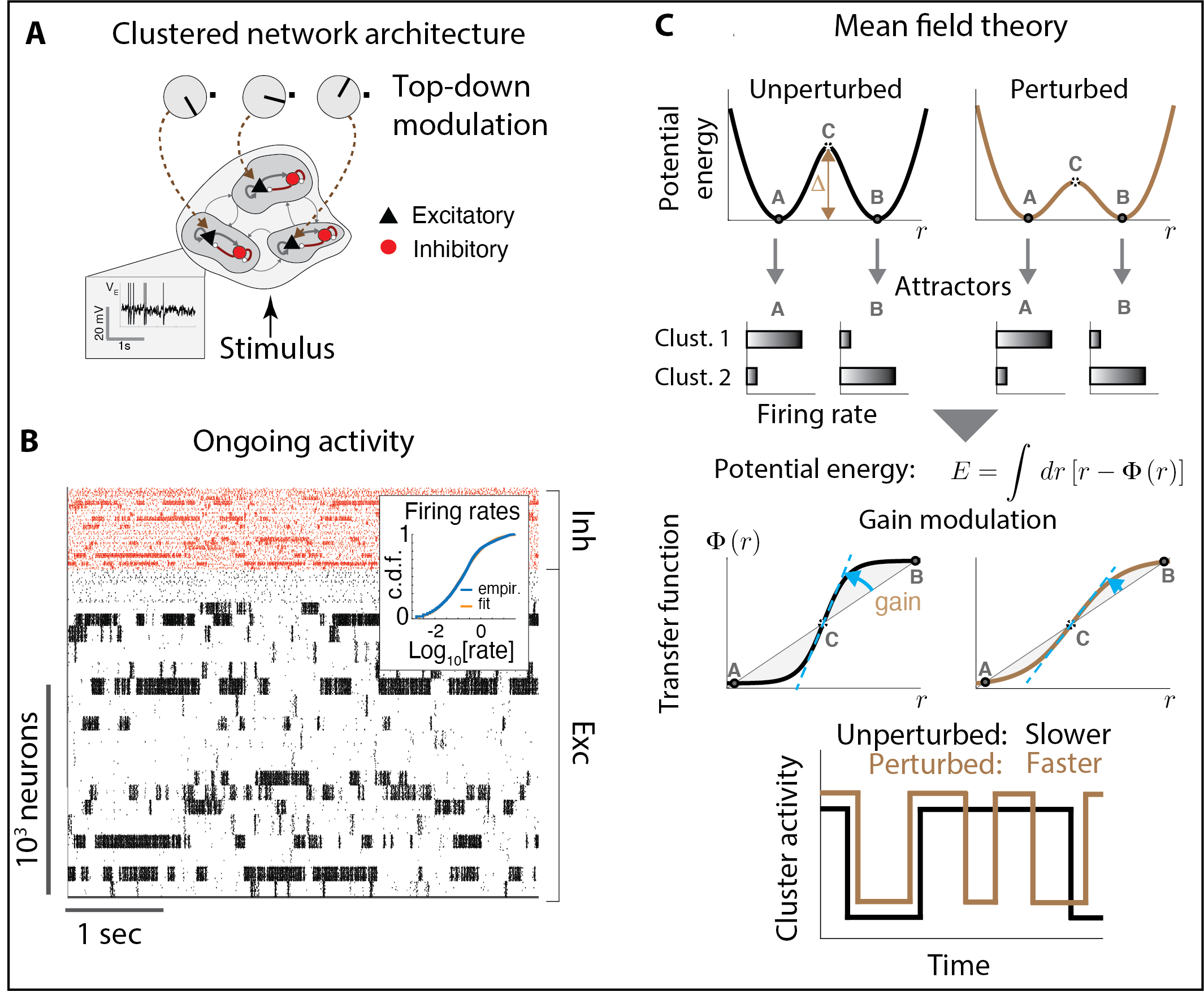}
\captionof{figure}{{\bf A)} Biologically plausible model of a cortical circuit generating metastable attractor dynamics. A recurrent network of E (black triangles) and I (red circles) spiking neurons arranged in clusters is delivered sensory stimuli and top-down modulation representing changes in cortical states. Inset: Membrane potential trace from a representative E neuron. {\bf  B)} Representative neural activity during ongoing periods (tick marks represent spike times of E (black) or I (red) neurons). The network activity unfolds through metastable attractors, each attractors corresponding to subsets of transiently active cluster. Inset: the cumulative distributions of single-cell firing rates is lognormal.  {\bf C)} The effect of perturbations on network dynamics in a two cluster network can be captured by a double-well potential (top). Potential wells represent two attractors where either cluster is active (A and B), separated by a barrier with height $\Delta$. Mean field theory links perturbation effects on the barrier height (top right, lower barrier) to changes in the intrinsic neuronal gain (center right, lower gain). Bottom: A perturbation that decreases barrier heights and gain leads to faster transitions between attractors. All panels are adapted from Fig. 2 and 3 \citep{wyrick2021state}. \label{figattractormeta}
}
\end{center}

Although in the models of memory and delay persistent activity discussed so far attractors are stable, noise originating from some stochastic process may destabilize them, generating spontaneous transitions between \lm{them}. The hallmark of these noise-driven metastable dynamics is a right-skewed \lm{distribution of attractor dwell times with an exponential tail}. The source of this stochasticity can be either external (i.e., incoming Poisson spike trains relaying neural activity from other brain areas) or intrinsically generated from within the local recurrent circuit. In the case of recurrent networks of rate units (Fig. \ref{figthree}D), metastable dynamics can be triggered by low-dimensional correlated variability arising from synaptic dynamics in a feedback loop between a cortical and a subcortical area \citep{recanatesi2022metastable}. In the case of a spiking circuit, intrinsic variability may arise from the asynchronous irregular spiking activity ubiquitously observed across many brain areas \citep{softky1993highly,shadlen1998variable}. The asynchronous irregular regime is a hallmark of neural circuits operating in a fluctuation-driven regime \citep{amit1997dynamics} and in balanced networks \citep{van1996chaos}. The stochasticity arising from irregular spiking activity can generate noise-driven transitions between stable attractors and represents a robust and parsimonious mechanism to explain the origin of metastable attractors \citep{deco2012neural,LitwinKumarDoiron2012,mazzucato2015dynamics,wyrick2021state}. This class of biologically plausible models is based on an architecture whereby excitatory and inhibitory neurons are arranged in neural clusters, or functional assemblies, where neurons belonging to the same cluster have potentiated synaptic couplings compared to neurons belonging to different clusters (Fig. A). During ongoing periods, in the absence of external stimulation, the network activity unfolds via a sequence of metastable attractors, where each attractor is defined by a subset of activated clusters (Fig. B). Using mean field theory, one can represent attractors as potential wells in an attractor landscape \citep{Mascaro1999,LitwinKumarDoiron2012,MattiaSanchezVives2012,wyrick2021state}. The attractor landscape can be calculated analytically using a Lyapunov function (in networks with symmetric synaptic couplings the only minima are point attractors (Fig. A), although in general more complex minima may arise \citep{brinkman2021metastable}). In a simplified network with two clusters (Fig. C), each well corresponds to a configuration where one cluster is active and the other is inactive. When the network activity dwells in the attractor represented by the left potential well, it may escape to the right potential well because of internally generated variability. This process will occur with a probability determined by the height $\Delta$ of the barrier separating the two wells: the higher the barrier, the less likely the transition \citep{hanggi1990reaction}. The mean first passage time for this process is captured by the Arrhenius law \citep{huang2017once}, which can be generalized to the more realistic case of transitions driven by colored noise \citep{stern2021reservoir}.

\subsection{Fast and slow fluctuations}
\lm{Given a multi-stable landscape with multiple fixed points, such as the one generated by clustered networks, there are specific requirements on the statistics of noise fluctuations in order to generate metastable dynamics. Within a mean field approach, we can approximate the on-off switches of a single cluster as driven by escape noise from the energy potential (Fig. C). There are at least three sources of noise fluctuations in this spiking circuit, contributing to the input current to the representative cluster: the fluctuations due to the finite size of neural clusters, the shot noise generated by the incoming Poisson spike trains from other recurrently coupled neurons, and the slow rate fluctuations due to the metastable dynamics of the other clusters in the network. Finite size effects can be modeled as white noise, whose amplitude is inversely proportional to the cluster size \citep{brunel2000dynamics,huang2017once}. This fast noise source is sufficient to account for the emergence of a slow metastable dynamics of the cluster's on-off switches, captured by Kramers theory of noise-induced escape in a bistable system \citep{hanggi1990reaction,huang2017once,LitwinKumarDoiron2012}. In this simple picture, increasing the cluster size has the dual effect of both reducing the noise amplitude and also increasing the potential barriers. However, the picture significantly changes if we account for the other two sources of noise, which are both colored. In the diffusion approximation for the spiking network, an incoming spike train filtered by exponential synapses gives rise to an Ornstein-Uhlenbeck process, whose color is set by the synaptic time constant and whose amplitude is proportional to recurrent coupling strength \citep{fourcaud2002dynamics}. Finally, the self-consistent rate fluctuations originating from the metastable dynamics of the other network clusters give rise to a colored noise whose slow timescale is proportional to the average switching time \citep{stern2021reservoir}. In order to deal with escape colored noise, we can replace the Kramers theory \citep{hanggi1990reaction,huang2017once,LitwinKumarDoiron2012}  with the universal colored noise approximation (UCNA) to the Fokker-Planck equation, which provides a good approximation of the escape process in this regime \citep{stern2021reservoir}. Here, both the strength of the noise amplitude and the barrier height are proportional to a product of the noise color and the recurrent coupling strength. This theory shows that the overall effect of colored noise is to slow down the average transition rate. Although UCNA has been applied to networks of rate units, a full account of the spiking network dynamics is still an open problem. In summary, in models with metastable attractors, we can find a separation of timescales between the fast fluctuations around a particular attractor, and the slow fluctuations involving switching between different attractors. Moreover, network simulations suggest that the details of the network architecture might play an important role in generating metastable dynamics over a large range of parameters. Whereas in networks with only excitatory but not inhibitory clusters the metastable regime might require fine tuning \citep{LitwinKumarDoiron2012,mazzucato2019expectation}, an architecture with coupled excitatory and inhibitory clusters allows metastable dynamics to persist over a wide range of synaptic coupling strength and cluster sizes \citep{rostami2020spiking,wyrick2021state}. This E/I clustered architecture is necessary to generate a heterogeneous distribution of timescales within the same circuit, as explained in Fig. \ref{figseven}E \citep{stern2021reservoir}. 
}

\subsection{Top-down modulation of network metastable dynamics} 
A central challenge in experimentally testing computational models of metastable attractors is that the reconstruction of the attractor landscape and the height of its potential wells relies on the knowledge of the network's structural connectivity including its coupling strength and connection probability, which is out of reach of current neurotechnology. However, one can use mean field theory methods to obtain an alternative formulation of the attractor landscape only involving quantities directly accessible to experimental observation \citep{Mascaro1999,mattia2013heterogeneous,mazzucato2019expectation,wyrick2021state}. The double potential well representing the two attractors can be directly mapped to the effective transfer function of a neural population representing a single cluster (Fig. C). Using this correspondence, one can map changes in the barrier height $\Delta$ separating metastable attractors to changes in the slope (or "gain") of the intrinsic transfer function estimated during ongoing periods. This map provides a direct relationship between changes in cluster activation timescale and single-cell gain modulation, which can be inferred from either intracellular or extracellular recordings \citep{lim2015inferring,wyrick2021state}. These computational tools can be used to test mechanistic hypothesis on the effect that circuit perturbations and manipulations exerts on a network metastable dynamics \citep{mazzucato2019expectation,wyrick2021state}. In particular, perturbations inducing steeper gain increase well depths and barrier heights, and thus increase the cluster timescale, and vice versa.

\lm{An exciting new direction aims to infer the dynamical system underlying the observed spiking activity directly from the data. Assuming that high dimensional neural activity is generated by a low-dimensional set of latent variables evolving according to Langevin dynamics, one can directly infer the effective energy potential underlying their dynamics from the data \citep{genkin2020moving,duncker2019learning}. A potential limitation of these new methods may be their reliance on low-dimensional latent manifolds, whereas the structure of neural patterns in motor areas during freely moving behavior seems to be high dimensional \citep{recanatesi2022metastable}.
}

\end{appendixbox}

\bibliography{bib}

\end{document}